\documentclass[aps,prb,superscriptaddress,amsmath,amssymb,onecolumn]{revtex4-1}

\usepackage{float, epsfig, subfigure, xcolor, amsmath, amsfonts, graphicx, gensymb}  
\usepackage[colorlinks=true, allcolors=blue]{hyperref}

\newcommand*{\citeref}{}
\DeclareRobustCommand*{\citeref}[1]{%
 \begingroup  
   \romannumeral-`\x 
   \setcitestyle{numbers}%
   \cite{#1}%
 \endgroup  
}  
  
\newcommand{\hw}{$\hbar \omega$}  
\newcommand{\Csxty}{$\text{C}_{60}$}  
\newcommand{\Ket}[1] {\ensuremath{\left| {#1} \right\rangle } }  
  
\begin{document}  

\title [FEL-induced damage in C and Si] {Various damage mechanisms in carbon and silicon materials \\ under femtosecond x-ray irradiation}  

\author{Nikita Medvedev\footnote{Corresponding author; E-mail: nikita.medvedev@fzu.cz}}  
\affiliation{Institute of Physics, Czech Academy of Sciences, Na Slovance 2, Prague 8, 18221, Czechia}  
\affiliation{Institute of Plasma Physics, Czech Academy of Sciences, Za Slovankou 3, Prague 8, 18200, Czechia}  

\author{Viktor Tkachenko}  
\affiliation{Institute for Laser and Optics, Hochschule Emden/Leer - University of Applied Sciences, Constantiaplatz 4, 26723 Emden, Germany} 
\affiliation{Center for Free-Electron Laser Science CFEL, Deutsches Elektronen-Synchrotron DESY, Notkestrasse 85,  22607 Hamburg, Germany}  

\author{Vladimir Lipp}  
\affiliation{Center for Free-Electron Laser Science CFEL, Deutsches Elektronen-Synchrotron DESY, Notkestrasse 85,  22607 Hamburg, Germany}  

\author{Zheng Li}  
\affiliation{Center for Free-Electron Laser Science CFEL, Deutsches Elektronen-Synchrotron DESY, Notkestrasse 85,  22607 Hamburg, Germany}  
\affiliation{Max Planck Institute for the Structure and Dynamics of Matter, 22761 Hamburg, Germany}  

\author{Beata Ziaja}  
\affiliation{Center for Free-Electron Laser Science CFEL, Deutsches Elektronen-Synchrotron DESY, Notkestrasse 85,  22607 Hamburg, Germany}  
\affiliation{Institute of Nuclear Physics, Polish Academy of Sciences, Radzikowskiego 152,  
31-342 Krak\'ow, Poland}  

\begin{abstract}  
We review the results of our research on damage mechanisms in materials irradiated with femtosecond free-electron-laser (FEL) pulses. They were obtained using our hybrid approach, XTANT (X-ray-induced Thermal And Nonthermal Transitions). Various damage mechanisms are discussed with respect to the pulse fluence and material properties on examples of diamond, amorphous carbon, \Csxty\ crystal, and silicon. We indicate conditions: producing thermal melting of targets as a result of electron-ion energy exchange; nonthermal phase transitions due to modification of the interatomic potential; Coulomb explosion due to accumulated net charge in finite-size systems; spallation or ablation at higher fluences due to detachment of sample fragments; and warm dense matter formation. Transient optical coefficients are compared with experimental data whenever available, proving the validity of our modeling approach. Predicted diffraction patterns can be compared with the results of ongoing or future FEL experiments. Limitations of our model and possible future directions of development are outlined. 
\end{abstract}
\date{\today}
\maketitle

\section{Introduction}
\label{Sec:Intro}

In the past decade, advances in extreme ultraviolet and x-ray laser science, in particular, the intense development of free electron laser (FEL) facilities~\cite{Pile2011, Altarelli2011, Allaria2012, Schreiber2015, Bostedt2016, Kang2017} have enabled time-resolved experiments, utilizing these unique radiation sources. A number of experiments probing x-ray induced dynamics at the femtosecond timescales have been performed, see, e.g.,~\cite{Gahl2008, Hau-Riege2012, Harmand2013, Tavella2017a}. 

FEL-induced ultrafast excitations of solids and phase transitions are nowadays crucial for understanding phenomena in various research fields, including solid-state physics~\cite{Krzywinski2007, Barty2008, Gaudin2013}, bio-physics~\cite{VanThor2015, Aquila2012, David2015}, physical chemistry~\cite{Polli2010, Li2013a}, plasma and warm dense matter (WDM)~\cite{Zastrau2012, Cho2016}. It is also invaluable for theorists to support their efforts to develop reliable simulation tools~\cite{Hau-Riege2011, Ziaja2013, Graziani2014, Jurek2016}.

Often advanced simulation tools for treating FEL-excited matter are based on the density functional theory molecular dynamics (DFT-MD) schemes, see, e.g.,~\cite{Vinko2010,Graziani2014,Bernardi2015,Dharma-wardana2016a}. Various improvements upon the standard DFT packages are being tested, such as incorporation of core-hole excitations~\cite{Kabeer2014}, or an attempt to reconcile bound and free electrons into a unique formalism~\cite{White2013}. However, important nonadiabatic effects governing electron-ion (electron-phonon) energy exchange are difficult to incorporate into \textit{ab-initio} simulations. The state of the art in the solid state community is so far an approximate treatment of electron-phonon scattering, valid only near the room temperature of atoms~\cite{Lin2008,Giustino2017}. In the \textit{ab-initio} femto-chemistry, advanced techniques allow for a treatment of nonadiabatic electron-ion coupling~\cite{Worth2004}. Attempts have been made to introduce similar methods into the solid-state models~\cite{McEniry2010}. But, at present, there is no standard methodology enabling to incorporate electron-ion interaction and the resulting energy exchange within \textit{ab-initio} approaches for highly-excited many-body systems.

Molecular dynamics methods with classical potentials or force fields are among the most commonly used numerical tools for large-scale simulations in the solid-state modeling, see, e.g.,~\cite{Zhigilei2006,Zhigilei2009,Eckhardt2013}. Utilization of classical interatomic potentials makes MD simulations very efficient and capable of treating large systems~\cite{Eckhardt2013,Wu2014}. However, such methods require predefined force fields, which applicability is limited to solids under low electronic excitation. A few attempts have been made to develop potentials depending on electronic temperature and/or number of excited electrons, e.g.,~\cite{Khakshouri2008,shokeen2010empirical}.

Classical Monte Carlo (MC) methods are often applied to simulate FEL-induced transport of electrons, photons or other particles in matter~\cite{Jenkins1988}. MC methods are suitable for simulating behavior of high-energy classical particles, whereas the propagation of low-energy particles involves strong quantum-mechanical effects which can hardly be treated. This limits the applicability of asymptotic-trajectory MC schemes.

Apart from the integral methods, differential methods are also used to treat systems with electronic excitations. Models operating with ensembles instead of individual particles can be derived from the Liouville equation. They are typically based on the single-particle distribution equations, such as kinetic equations ~\cite{Ziaja2008, Rethfeld2002}. Kinetic equations are capable of treating nonequilibrium evolution of solids. However, they are usually relying on a free-electron approximation or on a predefined band structure of a material ~\cite{Mueller2013}. In addition, the electron-ion coupling, defining energy exchange between the electronic and the atomic system, is one of the least known key parameters in these approaches. 

The next level of approximation is hydro- or thermodynamic methods~\cite{VanDriel1987,March1991,Inogamov2011}. The most commonly used is the Two-Temperature Model (TTM), which treats electrons and phonons as separate interacting subsystems in local equilibrium. Thermodynamic approaches are widely used due to their simplicity, but are limited to equilibrium conditions (see, e.g., review ~\citeref{Rethfeld2017}). TTM is inapplicable under non-equilibrium within electronic or atomic systems; in such cases, kinetic methods must be employed.  

Hybrid approaches are an actively developing field in numerical simulations. A hybrid approach combines two or more simulation techniques into a unified model. In such combinations, various approaches strengthen and complement each other, compensating mutually for their shortcomings, in particular, with respect to the overall computational efficiency. A proper combination of different models can alleviate limitations of each individual approach, thereby significantly extending the applicability of the combined model keeping the implementation simple enough. For example, a combination of the two-temperature model with molecular dynamics, TTM-MD~\cite{Ivanov2003}, was a major step that allowed to simulate experimental data on laser irradiation of solids with a high accuracy~\cite{ivanov2015experimental}.  Also, MC-MD combination allowed to treat accurately both atomic and electronic kinetics in case of plasmas~\cite{Murphy2014}. 

In this paper, we review our unique hybrid approach designed to treat solids under FEL irradiation. We demonstrate that such an approach is capable of capturing all the essential stages of the material evolution under femtosecond x-ray irradiation, starting from photo-induced nonequilibrium electron kinetics on femtosecond timescales, and progressing later towards atomic dynamics of thermal and nonthermal damage formation up to picosecond timescales. Thermal phase transition is a result of a kinetic energy exchange between hot electrons and atoms (e.g., electron-phonon coupling), occurring due to non-adiabatic coupling between the two systems. Nonthermal phase transition is a consequence of the changes of the potential energy surface for atoms due to the electronic excitations (e.g., nonthermal melting~\cite{Siders1999, Sundaram2002}).

In what follows, we present a comprehensive analysis of different damage channels in various carbon and silicon allotropes. Those materials are widely used in FEL-related optical elements, such as, e.g., x-ray mirrors. Their radiation tolerance plays a crucial role for this application. Our results indicate that different allotropes of carbon and silicon follow differing damage mechanisms under FEL irradiation.

With x-ray FELs, one can experimentally observe the evolution of diffraction patterns from irradiated samples with femtosecond resolution (e.g., Ref. \citeref{VanThor2015,Inoue2016}). Such a scheme is particularly interesting, because it can resolve the changing molecular structure of the sample, resulting from its progressing damage. Thus, throughout the paper, we present diffraction patterns calculated for each material and damage channels discussed. At the end, we discuss the applicability of the presented hybrid approach and possible pathways of its further development.

\section{Model}  
\label{Sec:XTANT}  
\subsection{Hybrid approach}  
\label{Sec:XTANT_scheme}  

\begin{figure*}[th!]  
\centering  
\includegraphics[width=0.94\linewidth,trim={20 190 20 0},clip]{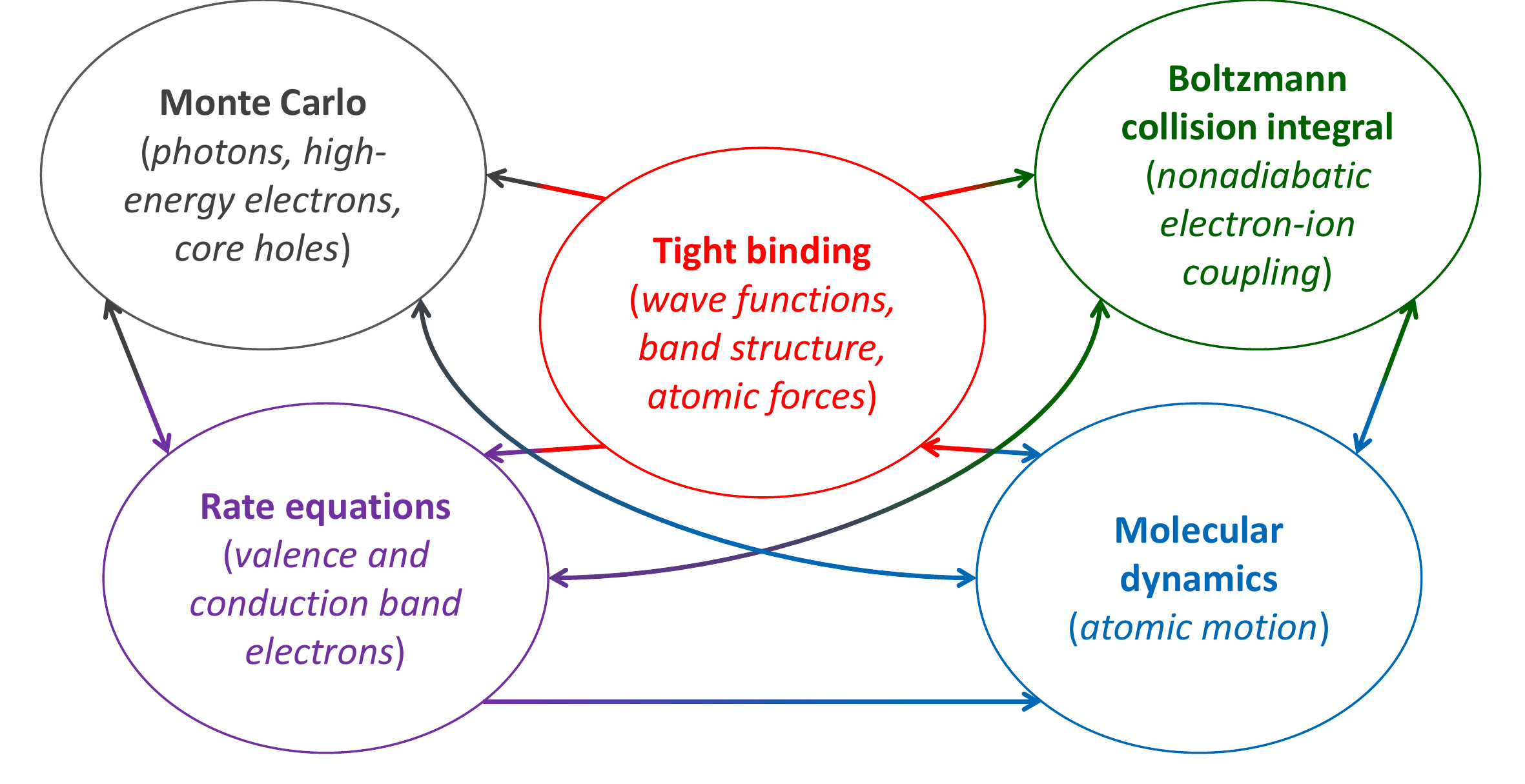}  
\caption{Schematics of the modular structure of hybrid code XTANT. Arrows indicate data flows between different modules of the program.}  
\label{fig:XTANT_modules}  
\end{figure*}

Our recently developed hybrid code XTANT (X-ray-induced Thermal And Nonthermal Transitions) is a combination of a few different schemes interconnected and executed in parallel. These approaches include: (a) Monte Carlo (MC) module tracing x-ray photon absorption, high-energy electrons and core holes kinetics; (b) a module describing the evolution of the low-energy electrons, using rate equations and thermodynamic modeling (in a similar manner to the two-temperature model, TTM); (c) the Boltzmann collision integral module following the nonadiabatic electron-ion energy exchange; (d) transferable tight binding (TB) model for calculations of the transient electronic band structure and atomic potential energy surface; and (e) Molecular Dynamics (MD) simulation tool to follow atomic motion. The scheme showing the interconnection of the most important modules is presented in Fig.~\ref{fig:XTANT_modules}. This diagram also indicates data flows between different modules at each time-step of the simulation. They will be described below in more detail.  

The chosen combination of approaches relies on the fact that the time-dependent electron distribution function, $f_{\rm e}(E_{i}, t)$, affects atomic motion. That is because transient electron distribution enters equations for the atomic potential energy surface (cf. Eq.\,(\ref{Eq:PotEn}))). To trace the evolution of the electron distribution function in time, we notice that the typical transient electron distribution after x-ray irradiation has the shape of the so-called 'bump on hot tail' \cite{Chapman2011}. It combines (i) a (nearly) thermalized  fraction of low-energy electrons within the valence band and the bottom of the conduction band, and (ii) a high-energy nonequilibrium tail containing a few highly energetic electrons left after photoionization and Auger decays. It has been proven theoretically that such a transient electron distribution is typical in various materials after an FEL pulse \cite{Faustlin2010,Chapman2011,Medvedev2011a,Hau-Riege2013}. Recently, predictions on the electron distribution function in FEL-irradiated aluminum from Ref. ~\citeref{Medvedev2011a}  were confirmed experimentally in Ref.~\citeref{Bisio2017}.

This particular shape of the electronic distribution function allows to simplify the model by combining two efficient approaches for low-energy and high-energy fractions of the electronic distribution function. It is used in the presented XTANT code as described in detail in Section \ref{Sec:High_en_electrons}.

Our hybrid model was specifically designed to follow the processes occurring in a solid target under irradiation with a femtosecond free-electron laser pulse. It is applicable in a broad range of photon energies: from extreme ultraviolet (XUV, photon energy above $\sim 30$ eV) to hard X--rays ($\sim 100$ keV photon energy)~\cite{Gaudin2013,Medvedev2013e,Medvedev2013f,Medvedev2015c,Medvedev2017}.  

\subsection{Creation and kinetics of high-energy electrons and core holes modeled with Monte Carlo method}  
\label{Sec:High_en_electrons}  

For the modeling of photon absorption, the kinetics of high-energy electrons, and Auger-decays of core holes, we apply an event-by-event individual-particle Monte-Carlo scheme \cite{Jacoboni1983,Jenkins1988,Akkerman1996}. 
For XUV and x-rays at fluences currently accessible at FELs, single photon absorption is the dominant interaction channel~\cite{Medvedev2010d}; other interaction channels such as elastic scattering, multiphoton absorption and inverse Bremsstrahlung (in XUV regime) are at least by two orders of magnitude less probable. An absorbed photon initiates the release of a high-energy photoelectron, and leaves a core-hole in an atom.

In the presented approach, the choice of whether the photoabsorption occurs by an excitation of an electron from the valence band or a deep shell is made, using the subshell photoabsorption cross sections taken from the EPDL97 database~\cite{Cullen1997}. Atomic cross sections for photoabsorption are applied, which are a good approximation for core shells. For the valence band, the photoabsorption cross section is obtained from the experimental optical coefficients~\cite{Palik1985}.

The released photoelectron has initial kinetic energy equal to the difference between the photon energy and its ionization potential, $I_p$. Ionization potentials for core shells are taken from EADL database~\cite{Perkins1991}, whereas for the valence or conduction band photoabsorption, the TB calculated transient energy levels are used (Eq.(\ref{Eq:TBdiagonalization}) in Section~\ref{Sec:TBMD}).  

High-energy electrons, i.e., those populating states at energies above a certain energy threshold, are treated as classical individual particles within the MC routine. In the presented calculations the threshold is chosen to be $E_{\rm cut} = 10$ eV, counted from the bottom of the conduction band. Electrons with lower energies are attributed to the 'low-energy' domain (see Section~\ref{Sec:Low_En_electrons})~\cite{Medvedev2015}. Alternatively, the threshold can be chosen as equal to the uppermost energy level of the conduction band produced by the TB calculations. Influence of the cut-off value on the results was analyzed in Ref.~\citeref{Medvedev2013f}, showing almost no effect on the results, if varied by a few eV around the $10$ eV value.

Each high energy electron can scatter inelastically on the atomic core-shells, if its energy is higher than the respective binding energy, $E_{\rm e} > I_p$; otherwise, only the scattering on the valence-band electrons is possible.  
The electron mean free paths are estimated via the scattering cross sections, $\sigma_i(E_{\rm e})$, within the first Born approximation in terms of the complex dielectric function (CDF), $\epsilon(\omega_{\rm e},q)$~\cite{Ritchie1977,Akkerman1996,Medvedev2015a}:  
\begin{eqnarray}{}
\frac{d \sigma_i(E_{\rm e}, \hbar \omega_{\rm e})}{d (\hbar \omega_{\rm e})} = \frac{2 e^2 n_{\rm a}}{\pi \hbar^2 v^2} \int_{q_-}^{q_+} \frac{dq}{q}{\rm Im} \left( \frac{-1}{\epsilon(\omega_{\rm e},q)} \right)  \,  
\label{Eq:cross-section}  
\end{eqnarray}  
with $ q_{\pm} = \sqrt{2 m_{\rm e}/\hbar^2}\left( \sqrt{E_{\rm e}} \pm \sqrt{E_{\rm e} - \hbar \omega_{\rm e}} \right)$ \cite{Akkerman1996}. Here the cross section also depends on the energy \hw$_{\rm e}$ gained by the secondary electron in a collision, and is integrated over the transferred momentum $q$; $e$ denotes the electron charge; $m_e$ is the free electron mass; $n_{\rm a}$ is the atomic density; $\hbar$ is the Planck constant; and $v$ is the incident electron velocity corresponding to the kinetic energy $E_{\rm e}$.
The CDF is parameterized for each shell of each element; all the parameters and accuracy checks of the cross sections can be found in Ref.~\citeref{Medvedev2015a}. The calculated electron inelastic mean free paths showed a very good agreement within a few percent with the NIST database~\cite{Powell1999} and available experimental data for electron energies above $\sim 50$ eV.  

When a high-energy electron collides with the valence or deep shell electrons, the probabilities for these collisions are estimated using the scattering cross sections from Eq.(\ref{Eq:cross-section}). The initial energy of the secondary electron emitted during an inelastic collision is calculated from the energy conservation, i.e., it equals to the difference between the energy lost by the incident electron and the binding energy of the level, from which this electron is being ionized.
If the energy of any electron falls below the $E_{cut}$, this electron is removed from the MC domain and added into the low-energy domain. Detailed description of this inter-domain energy and particle exchange is presented in section \ref{Sec:Low_En_electrons}.  

The elastic scattering of electrons on ions is modeled using Mott's cross section with modified Moliere screening parameter~\cite{Jenkins1988}. However, this interaction channel produces only a minor energy loss for electrons (and correspondingly negligible energy increase for atoms)~\cite{Lorazo2006,Medvedev2013e,Medvedev2013f}. Most of the energy transfer to the ions is produced by low-energy electrons via nonadiabatic coupling, as will be described in detail (Section~\ref{Sec:Low_En_electrons}).  

After a photoionization or impact ionization of a core shell, a hole is left behind. Predominantly, for light elements, this hole will decay via Auger processes~\cite{Keski-Rahkonen1974}. We use the Poissonian probability distribution to model Auger decays. The characteristic hole decay times are taken from the EADL database \cite{Perkins1991}. When a core hole relaxes during an Auger process, one electron is promoted from an upper shell or the valence band into high energy states of the conduction band, leaving another hole. In case of the valence band, the energy level from which the Auger electron is chosen randomly among all currently populated levels $E_{i}$. The Auger electron receives the excess energy and is then treated in the same way as other secondary electrons.

The MC tracing of photons, electrons, and core holes is split into time steps $dt$ equal to the molecular dynamics time steps. The number of iterations of the MC subroutine is proportional to the photon energy. This is necessary in order to obtain sufficient statistics to follow accurately small numbers of high-energy photo-electrons. E.g., for intermediate photon energies of a few keV, each time step of MC is iterated for more than 30000 times for reliable statistics. Trajectories of all electrons are propagated simultaneously during each iteration. The calculated electron distributions are then statistically averaged. 

As we assume homogeneous excitation, we neglect a contribution of electron transport and heat diffusion to the overall electron  kinetics. This can be justified as an x-ray irradiation homogeneously heats up the sample down to a few micrometers depth within the laser spot of typically a few microns size. The approximation of homogeneity allows us to use periodic boundary conditions, but excludes from our considerations the thin near-surface layer, from which the high-energy electrons could escape outside the material. To account for the effects of electron emission whenever needed, we artificially remove an electron from the simulation box after a certain number of collisions, introducing charge non-neutrality. This charge non-neutrality then contributes to additional Coulomb forces acting on atoms (Section~\ref{Sec:TBMD}).

\subsection{Low-energy electrons and nonadiabatic coupling modeled with rate equations including Boltzmann collision integral}  
\label{Sec:Low_En_electrons}  

At each time step, we track how the electrons are distributed between the high- and low-energy domains. The total number of low-energy electrons, $N_{\rm e}^{\rm low}$, and their energy, $E_{\rm e}^{\rm low}$, are calculated knowing how many electrons were excited to the high-energy domain by the incoming photons from the laser pulse, $N_{\rm ph}$, by the secondary electron collisions, $N_{\rm imp}$, by Auger-decays of core-shell holes, $N_{A}$, and by electron emission in case of thin films, $N_{\rm em}$. Electrons, which are excited to the energy levels above the cut-off energy, are transferred to the high-energy domain and treated with the MC algorithm (section \ref{Sec:High_en_electrons}). Vice versa, when an electron from the high-energy fraction loses its energy below the cut-off energy, it joins the low-energy domain. Thus, the total number of low energy electrons is calculated as:
\begin{eqnarray}{}
\label{Eq:Ne}
N_{\rm e}^{\rm low}(t+dt) = N_{\rm e}^{\rm low}(t) + {N}_{\rm e}^{\rm high}(t) 
- N_{\rm ph}(t) - N_{\rm imp}(t) - N_{A}(t) - N_{\rm em}, \  
\end{eqnarray}
%
where ${N}_{\rm e}^{\rm high}(t)$ denotes the fraction of high-energy electrons that fell into or were excited off the low-energy domain, as counted within the MC module. The total number of electrons in high- and low-energy domains, including those originating from core-shell ionizations and electron emissions, is conserved.  

The total energy of low-energy electrons, $E_{\rm e}^{\rm low}$, is calculated with an equation similar to (\ref{Eq:Ne}), in the following way:  
\begin{eqnarray}{}
E_{\rm e}^{\rm low}(t+dt) = E_{\rm e}^{\rm low}(t) + {E}_{\rm e}^{\rm high}(t) 
+ E_{\rm imp}(t) + E_{A}(t) - \delta E_{el-ion}, \  
\label{Eq:Ee}  
\end{eqnarray}
%
where $E_{\rm e}^{\rm low}$ is the total energy of low-energy electrons; ${E}_{\rm e}^{\rm high}(t)$ is the energy brought in or out by high-energy electrons that fell into or jumped off the low-energy domain (e.g by photoabsorption); $E_{\rm imp}$ is the energy delivered during the impact ionization events by the high-energy electrons; $E_{A}$ is the energy delivered by Auger-decays of core-shell holes that involve valence or conduction band electrons; and $\delta E_{el-ion}$ is the energy transferred to (or from) ions. The electron-ion energy exchange is calculated using the Boltzmann collision integral, $I_{i,j}^{\rm e-at}$:  
\begin{widetext}  
\begin{eqnarray}{}  
\label{Eq:Boltzmann}  
I^{\rm e-at}_{i,j} = w_{i,j}  
\begin{cases}  
f_{\rm e}(E_i)(2 - f_{\rm e}(E_j)) - f_{\rm e}(E_j)(2 - f_{\rm e}(E_i))G_{\rm at}(E_i - E_j) \ , {\rm for} \ i > j , \\  
f_{\rm e}(E_i)(2 - f_{\rm e}(E_j))G_{\rm at}(E_j - E_i) - f_{\rm e}(E_j)(2 - f_{\rm e}(E_i)) \ , {\rm for} \ i < j ,\  
\end{cases}  
\label{Fin_coll_int}  
\end{eqnarray}  
\end{widetext}  
where $w_{i,j}$ is the rate for an electron transition between the energy levels $i$ and $j$; $f_{\rm e}(E_i)$ is a transient electron distribution function; and $G_{at}(E)$ is the integrated Maxwellian function for atoms~\cite{Medvedev2015c}. One could, in principle, use the transient atomic distribution function obtained from the MD simulations, but due to relatively small number of atoms in the supercell, the fluctuations do not allow us to obtain a smooth function. This, in turn, introduces a numerical dependence of the energy transfer rate on the number of atoms. For this reason we approximate the atomic distribution as an equilibrium one, using a Maxwellian function.  

The following expression suitable for finite-difference implementation for the transition probability is used~\cite{Medvedev2017}:  
\begin{eqnarray}{}  
\label{Eq:Landau}  
w_{i,j} = \left| \left( \langle i(t) | j(t+\delta t) \rangle - \langle i(t +\delta t) | j(t) \rangle \right)/2 \right|^2\frac{1}{\delta t},  
\end{eqnarray}  

Note that this expression for the transition rate only reduces to the Fermi's Golden Rule in case of periodic atomic vibrations (phonons), and at long timescales in comparison to a duration of each individual act of scattering~\cite{Medvedev2017}. None of these assumptions holds true at subpicosecond timescales in case of irradiation with femtosecond laser pulses; thus, more general expression, Eq.(\ref{Eq:Landau}), has to be used in this case.  

The knowledge of the collision integral $I_{i,j}^{\rm e-at}$ allows one to evaluate the energy flux between the electrons and ions at each time step in Eq.(\ref{Eq:Ee}):  
\begin{eqnarray}{}  
\label{Eq:Heat_rate}  
\delta E_{el-ion} = \sum_{i,j} I_{i,j}^{\rm e-at} \cdot E_{i}  
\end{eqnarray}  
where the summation runs over all the electronic orbitals for transitions between each pair of levels~\cite{Medvedev2017}.

Knowing the total energy and number of the low-energy electrons at each time-step, we can estimate their temperature and chemical potential. They are calculated from the $0^{\rm th}$ and the $1^{\rm st}$ moments of the Fermi distribution function by solving the inverse problem ~\cite{Landau1975}:  
\begin{eqnarray}{}  
N_{\rm e}^{\rm low} = \sum_{E_{\rm min}}^{E_{\rm cut}} f_{\rm e}(E_{i}) = \sum_{E_{\rm min}}^{E_{\rm cut}} \frac{2}{1+\exp\left((E_{i} - \mu)/T_{\rm e}\right)}  ,\ \nonumber \\  
E_{\rm e}^{\rm low} = \sum_{E_{\rm min}}^{E_{\rm cut}} E_{i} \cdot f_{\rm e}(E_{i}) = \sum_{E_{\rm min}}^{E_{\rm cut}} \frac{2E_{i}}{1+\exp\left((E_{i} - \mu)/T_{\rm e} \right)}  , \  
\label{Fermimoments}  
\end{eqnarray}  
where the summations include all energy levels, $E_{i}$, corresponding to the current band structure of the material (that evolves in time), and $E_{\rm min}$ is the lowest energy level of the valence band. These energy levels are calculated with the tight binding method (section~\ref{Sec:TBMD}, Eq.(\ref{Eq:TBdiagonalization})). The factor $2$ in the Fermi-distribution function $f_{\rm e}(E_{i})$ accounts for the electron spin; $\mu$ is the transient chemical potential of the electrons and $T_{\rm e}$ is their temperature (in energy units). We solve the system of equations (\ref{Fermimoments}) for the known values of $N_{\rm e}^{low}$ and $E_{\rm e}^{low}$ at each time step by the bisection method. In this way, we find the transient values of $\mu$ and $T_{\rm e}$~\cite{Medvedev2013e,Medvedev2013f}.  

Note that, in principle, the model does not require the low-energy electron distribution to necessarily obey the equilibrium Fermi function. It can work for any transient nonequilibrium distribution function, such as provided by the Boltzmann equation~\cite{Rethfeld2002,Ziaja2005}. For example, we tested the possibility of independent distributions of holes in the valence and of electrons in the conduction band~\cite{Tkachenko2017}. For this case, the system of equations (\ref{Fermimoments}) can be written for each band independently, and each band can independently exchange the energy with the high-energy electrons in the MC domain. The work on this scheme is in progress, and will not be discussed in the current paper. Correspondingly, below we assume instant low-energy electron thermalization within the entire low-energy domain (Eqs.(\ref{Fermimoments})), which significantly simplifies the calculations.  

\subsection{Transferable tight binding molecular dynamics}  
\label{Sec:TBMD}  

To trace material modifications on the level of both electronic and atomic processes, we employ tight-binding molecular dynamics \cite{Xu1992,Kwon1994,HaraldO.Jeschke2000}. This method relies upon transferable tight binding Hamiltonian to evaluate electronic energy levels (band structure) and the interatomic potential energy surface. Transferable TB means that for given atomic species the parameterizations of the hopping integrals and the repulsive potential are constructed to reproduce many material phases.

The transferable TB Hamiltonian is written as follows \cite{Xu1992,Kwon1994,HaraldO.Jeschke2000}:  
\begin{eqnarray}{}  
H = H_{\text{TB}} + E_{\text{rep}}(\{ {\bf r}_{i j} \}) \ , \ H_{\text{TB}}=\sum_{i j \eta \nu} H_{i \eta j \nu} \ , \ \nonumber \\  
H_{i \eta j \nu} = \epsilon_{i \eta} \delta_{i j} \delta_{\eta \nu} + t_{i j}^{\eta \nu}(1 - \delta_{i j}) \ . \  
\label{Eq:TBHamiltonian}
\end{eqnarray}
where $E_{\rm rep}(\{ {\bf r}_{i j}\})$ is the repulsive part describing the effective repulsion of atomic cores and $H_{\text{TB}}$ is the attractive part calculated with the tight binding Hamiltonian. The TB part is constructed of the on-site energies, $\epsilon_{i \eta}$, and the pairwise overlap integrals, $t_{i j}^{\eta \nu}$, within the sp$^3$ basis set. 
Those are parameterized functions, which for the case of carbon-based materials can be found in Ref.\citeref{Xu1992}, and for silicon in Ref.\citeref{Kwon1994}.  

In the case of an orthogonal Hamiltonain parameterization, the electron energy levels are obtained by a direct diagonalization:  
\begin{eqnarray}{}
 E_{i} = \langle i | H(\{R_{\rm at}(t)\}) | i \rangle. \label{Eq:TBdiagonalization}
\end{eqnarray}

The potential energy surface $\Phi(\{r_{ji}\},t)$, needed as an input to the equations of motion for atoms, can be derived from Eq.(\ref{Eq:TBHamiltonian}) within the Born-Oppenheimer (BO) approximation from the Hellman-Feynman theorem as follows:  
\begin{eqnarray}{}  
\Phi(\{ r_{ij}(t)\}, t) = \sum_{\rm i} f_e(E_{i}, t) E_{i} + E_{\rm rep}( \{ r_{ij} \} )  \ .  
\label{Eq:PotEn}  
\end{eqnarray}  
Here, $f_{\rm e}(E_{i}, t)$ is the transient electron distribution function indicating fractional electron population numbers on the transient energy levels $E_{i}$ given by Eq.(\ref{Eq:TBdiagonalization}). 

Additional terms beyond the BO approximation result from the energy transferred from the electron in nonadiabatic transitions between the energy levels mediated by the atomic displacements~\cite{Medvedev2017}. This transferred energy is then distributed among all the atoms in the simulation box by the appropriate velocity scaling. The calculation of such energy transfer was described in the section \ref{Sec:Low_En_electrons}.  

The typical laser spot radius for an FEL laser is of the order of a few micrometers, and a photon penetration depth may be also on a few micron scale. This volume corresponds to at least a few billion atoms. Thus, we can choose only a small simulation box (supercell) inside the laser spot with a size much smaller than the laser spot, and apply periodic boundary conditions. The periodic boundary conditions in XTANT can be introduced in two ways. First, we can keep the super-cell vectors constant during the simulation, modeling the NVE ensemble (i.e. constant volume simulation). Alternatively, within the Parrinello-Rahman method \cite{Parrinello1980}, we can account for the changing geometry of the supercell. It is traced via additional variables entering the Lagrangian of motion~\cite{Parrinello1980,HaraldO.Jeschke2000}, for the NPH ensemble (i.e. assuming constant pressure simulation). 
We use the velocity Verlet algorithm for propagating atomic coordinates and velocities in time~\cite{Medvedev2013e}. The applied time-step is usually 0.01 fs, ensuring a stable numerical scheme~\cite{Medvedev2017}.

For the van der Waals forces acting on carbon atoms in case of \Csxty\ crystal or graphite, we employ additional semiempirical Lennard-Jones (6-12) potential, softly cut at short and large distances~\cite{Toufarova2017}. Soft cut-off at short distances ensures that it does not overlap with the short-range forces treated within the TB approach.  

In case of charge non-neutrality which may occur in thin films after electron emission (see section \ref{Sec:High_en_electrons}), unbalanced positive charge is then accounted for as an additional fractional charge equally distributed among all the atoms in the simulation box~\cite{Toufarova2017}. It produces additional long-range Coulomb potential, which was not included in TB.

\subsection{Data analysis based on optical properties, autocorrelations and diffraction patterns}  
\label{Sec:Data_analysis}  

Throughout the paper, we define the absorption dose as the absorbed energy per atom on the depth equal to the attenuation length of the considered photon:  
\begin{eqnarray}{}
D_{\rm abs} = F(1-R)(1-\exp(-d/\lambda))/(n_{\rm at} d) = 
F(1-R)(1-\exp(-1))/(n_{\rm at} \lambda)  
\label{Eq:Dose_fluence}  
\end{eqnarray}
where $D_{\rm abs}$ is the absorbed dose (eV/atom); $F$ is the incoming fluence into the material; $R$ is the reflectivity of the sample at the wavelength of the incoming pulse; $d$ is the material thickness, assumed to be equal to the photon attenuation length under normal incidence $\lambda$; $n_{\rm at}$ is the atomic density. For comparison with experiments, one may use Eq.(\ref{Eq:Dose_fluence}) to evaluate the incoming fluence from a given dose. Such a connection assumes linear photoabsorption (without multiphoton effects), which is generally a good approximation for photon energies above a few tens of eV, typical for FELs.


Knowing atomic positions and the supercell vectors from the MD module, one may obtain powder diffraction patterns with available software. We calculate the patterns with help of the Mercury software~\cite{Macrae2008}, renormalizing them to the area under the peaks instead of the highest peak height.

We obtain the vibration spectrum from the time evolution of the atomic velocities $\vec{v}_k(t)$ in excited solids, with consideration of the large amplitude motion~\cite{pra04:11056,kir12:6608,ven15:8080}.  
%
\begin{eqnarray}  
 \label{Eq:spectrum}  
 I(\tau,\omega)= \mathcal{F}_t[I(\tau,t)](\omega) 
 =\mathcal{F}_t\left\{ \frac{1}{N_{at}}\sum_{k}^{N_{at}}\left\langle \vec{v}_k(\tau)\cdot\vec{v}_k(\tau +t)\exp(-\alpha t^2) \right\rangle \right\}(\omega)  
\,  
\end{eqnarray}  
%
where $\mathcal{F}_t$ denotes the Fourier transform; $N_{at}$ is the number of atoms, and the parameter $\alpha$ is chosen to suppress the autocorrelation function within 200 fs to effectively select the vibrational modes that are present within a given time interval $\tau$ to $\sim\tau+200$fs.
In this way, the temporal evolution of vibrational modes can be revealed by their frequencies and amplitudes. In particular, it allows to trace disappearance of a harmonicity in the atomic motion, indicating loss of the structure in the crystal, i.e., phase transition to a disordered phase. 

Transient optical properties of materials, such as reflectivity or transmissivity, can also be measured in pump-probe experiments with a time resolution down to 10 fs~\cite{Tavella2017a,Maltezopoulos2008,Harmand2013}. Thus, along with diffraction patterns the latter can be used  
for comparison between the experimental and simulation results. Within the linear response theory, optical properties are defined by the complex dielectric function (CDF). The random phase approximation (RPA) provides the following expression for the dielectric function (the Lindhard formula)~\cite{Ehrenreich,Trani2005}:  
\begin{equation}
\varepsilon^{\alpha\beta}(\omega)= 1 + {{{e}^2\,{\hbar}^2}\over \,{m_e}^2\Omega\,\epsilon_0}\sum_{\eta\nu}{{F_{\eta\nu}}\over \,E_{\eta\nu}^2}\,{f_{\nu}-f_{\eta}\over \hbar w-E_{\eta\nu}+i\,{\gamma}}.  
\label{Eq:rpa}  
\end{equation}  
Here, $E_{\eta\nu} = E_{\nu} - E_{\eta}$ is the transition energy between two eigenstates $\Ket{\eta}$ and $\Ket{\nu}$;  
$f_{\eta}$ and $f_{\nu}$ are the transient occupation numbers of the corresponding states as defined above; $F_{\eta\nu}$ are the diagonal elements of the oscillator strength matrix~\cite{Trani2005,Tkachenko2016}; $\Omega$ is the volume of the supercell; and $\epsilon_0$ is the vacuum permittivity in SI units.

A particular choice of the (small) parameter $\gamma$ does not affect the results beyond the broadening of peaks in the CDF~\cite{Klingshirn}. The dependence of the results on $\gamma$ was investigated in detail in Ref.~\citeref{Medvedev2017a}.  

The real, $\varepsilon'$, and imaginary, $\varepsilon''$, parts of the CDF define the components $n$ and $k$ of a complex index of refraction $\widetilde{n}(\omega) = n + i k$ by relations:  
\begin{eqnarray}  
n^2 &= {1 \over {2}}\left(\sqrt{\varepsilon'^{\,2} + \varepsilon''^{\,2}} + \varepsilon'\right), \ \nonumber \\  
k^2 &= {1 \over {2}}\left(\sqrt{\varepsilon'^{\,2} + \varepsilon''^{\,2}} - \varepsilon'\right).  
\label{refraction_index}  
\end{eqnarray}  
Using Fresnel's and Snell's laws, the reflectivity coefficient can be expressed as follows~\cite{Yeh2005}:  
\begin{equation}  
R = \left|{{\cos{\theta} - \sqrt{{\widetilde{n}}^2 - \sin^2{\theta}}}\over \,{\cos{\theta} + \sqrt{{\widetilde{n}}^2 - \sin^2{\theta}}}}\right|^2,  
\label{reflectivity}  
\end{equation}  
where $\theta$ is the angle of incidence of the probe pulse.  

The transmission coefficient of the material also depends on the material thickness $d$ and the wavelength of the incident probe pulse $\lambda$. In the case of a bulk or a thick layer of a material and ultrashort probe pulse, we assume the first ray propagation with no interference effects included from multiple reflections on the material boundaries~\cite{Yeh2005,Tkachenko2016}:  
\begin{equation}  
T = {\left|{4\cos{\theta}\sqrt{{\widetilde{n}}^2 - \sin^2{\theta}}\cdot e^{-i\,{{2\pi d}\over \,{\lambda}}\sqrt{{\widetilde{n}}^2 - \sin^2{\theta}}}\over \,(\cos{\theta} + \sqrt{{\widetilde{n}}^2 - \sin^2{\theta}})^2}\right|^2},  
\label{trasmittivity}  
\end{equation}  
The absorption coefficient can then be obtained from the normalization condition:  
\begin{equation}  
A = 1 - T - R.  
\label{absorption}  
\end{equation}  
%

\section{Results}  

\subsection{Low fluence}  

\subsubsection{Nonthermal graphitization of diamond}  
\label{Sec:Nonthermal_diamond}  
 
In a series of papers, we modeled diamond under femtosecond FEL irradiation in a wide range of photon energies, using XTANT code described above~\cite{Medvedev2013c, Gaudin2013, Medvedev2013f, Medvedev2015, Tavella2017a}. We showed that at the absorbed doses above the damage threshold of $\sim 0.7-0.75$ eV/atom, diamond undergoes a nonthermal solid-to-solid phase transition into graphite phase on an ultrashort timescale of $\sim~150$ fs. An average dose of above 0.7 eV per atom in diamond leads to the excitation of over 1.5\% of electrons from the bonding states of the valence to the antibonding states of the conduction band. This is sufficient to trigger the graphitization.

Non-thermal graphitization of diamond proceeds in the following steps~\cite{Medvedev2013c}:

(i) Initial electronic excitation occurs during the FEL pulse. In case of x-ray pulses, photoelectrons relax to the bottom of the conduction band within a few up to a few-tens of femtoseconds (depending on the photon energy, see, e.g., ~\cite{Medvedev2015a}) via collisional processes and Auger recombinations of K-shell holes. During this step, low-energy electrons in the valence and the bottom of the conduction band receive energy from the high-energy electrons, and are starting to exchange it with the lattice.  

(ii) Electronic excitation triggers a band gap collapse (see, e.g.,~\cite{Medvedev2013c}). It occurs within $\sim 50$ fs (for soft x-rays) after the pulse maximum of the FEL pulse at the time instant when the density of conduction band electrons overcomes the threshold value of $\sim 1.5 \%$, as mentioned above. This is accompanied by the interatomic sp$^3$ bonds breaking~\cite{Medvedev2013c}. For higher photon energies, the electron cascades last longer, thereby delaying all the ensuing processes~\cite{Medvedev2015}.  

(iii) These processes are followed by the atomic relocation (occurring at $\sim 150-200$ fs), which significantly changes the material properties: from insulating diamond to semi-metallic graphite. The electronic density in the conduction band further increases, leading to the final irreversible atomic relocation~\cite{Medvedev2013c}. In this rearrangement, atoms settle at the new positions corresponding to overdense graphite.  
\begin{figure*}  
 \includegraphics[width=0.9\textwidth, trim = {30 300 20 20}]{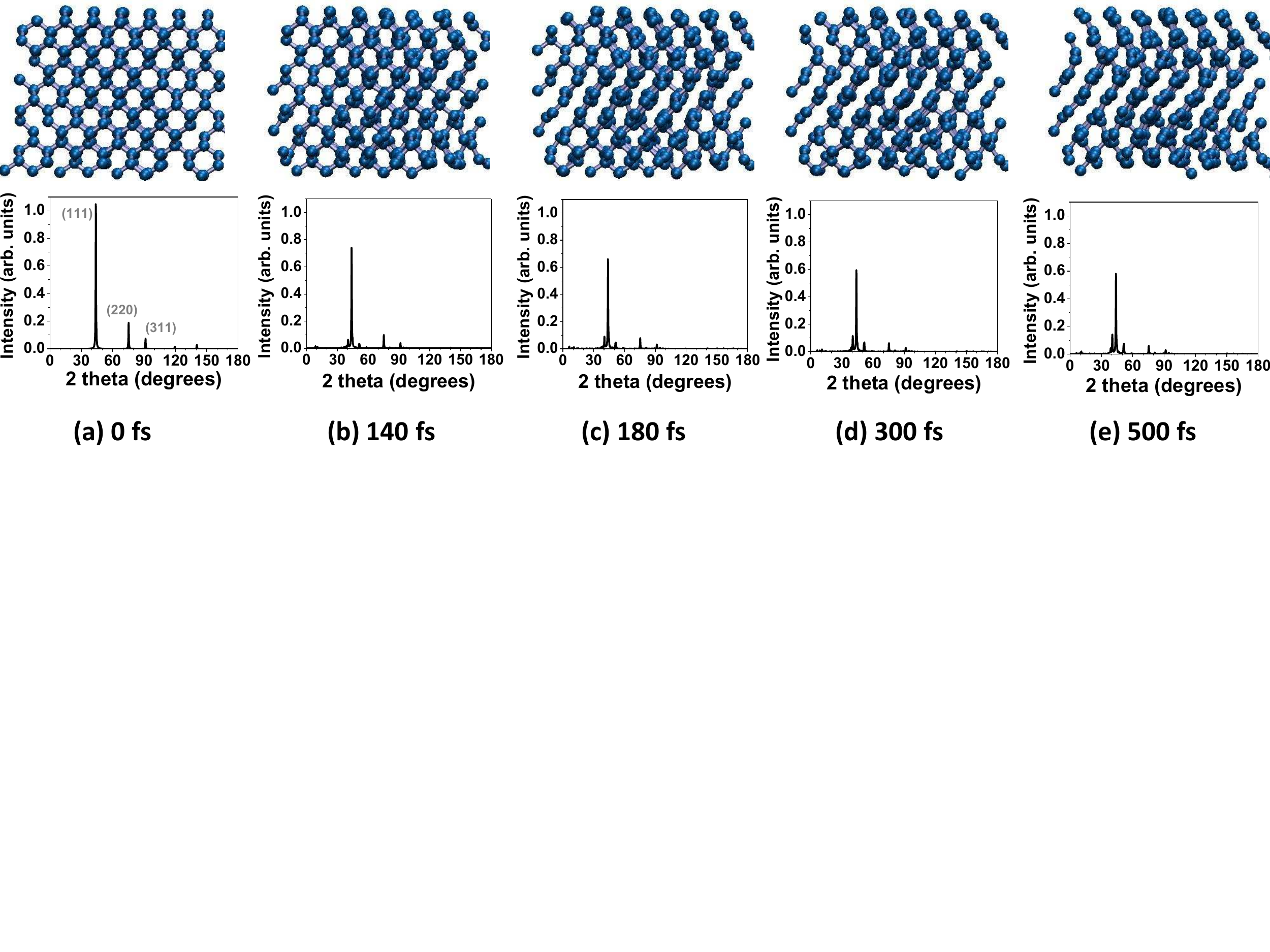}  
 \caption{\label{fig:Diamond_diffractions}  
 (Top raw) Snapshots of diamond during graphitization at different time instants after the FEL pulse irradiation corresponding to the absorbed dose of 0.8 eV/atom, photon energy of 47.4 eV, and FWHM pulse duration of 52.5 fs; reproduced from Ref.~\citeref{Tavella2017a}. (Bottom raw) Powder diffraction patterns for the corresponding atomic structures obtained with x-ray photons of $1.54$ \AA.}  
\end{figure*}

\begin{figure*}  
 \includegraphics[width=0.9\textwidth, trim = {0 10 0 0}]{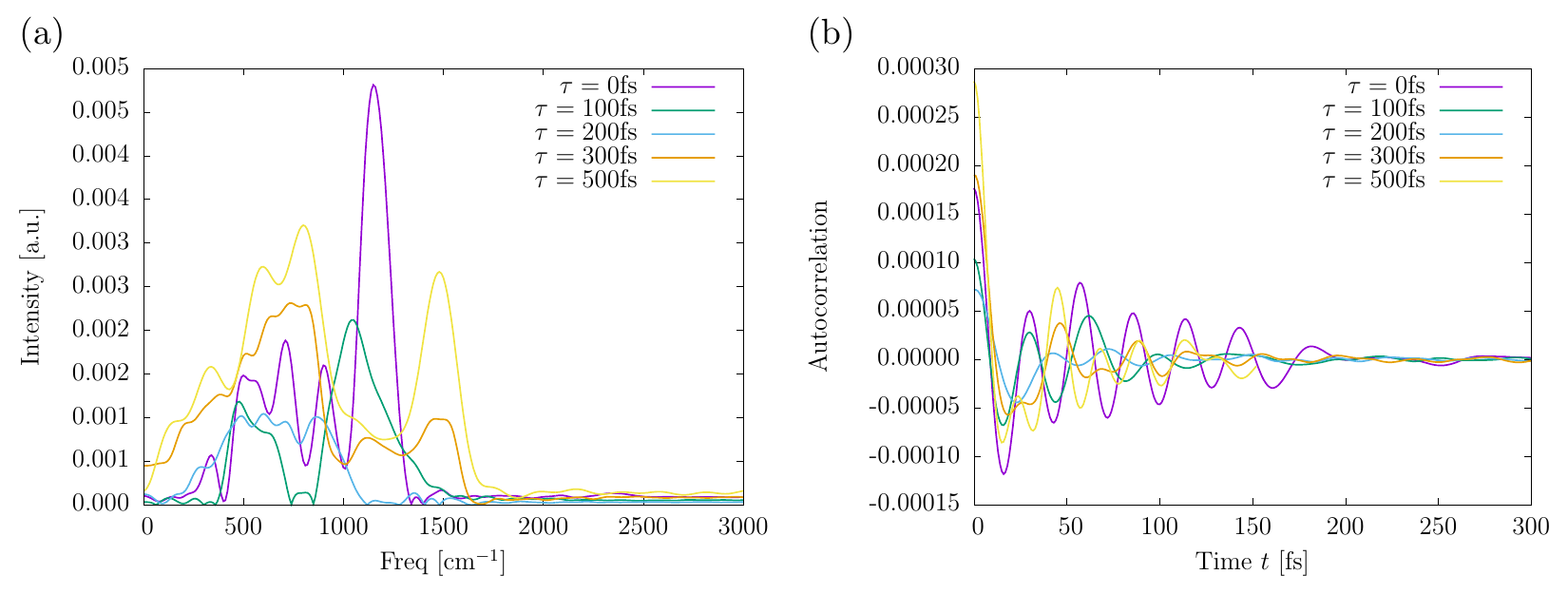}  
 \caption{\label{fig:Diamond_0.7_autocorr}  
 Calculated vibrational spectrum from the autocorrelation function of swarm of trajectories:  
 (a) vibrational spectrum for various time delays $\tau$ after the pump pulse; 
 (b) autocorrelation function for various time delays $\tau$, where the time $t$ is defined in Eq.~(\ref{Eq:spectrum})  
 }  
\end{figure*}

\begin{figure}  
 \includegraphics[width=0.45\textwidth]{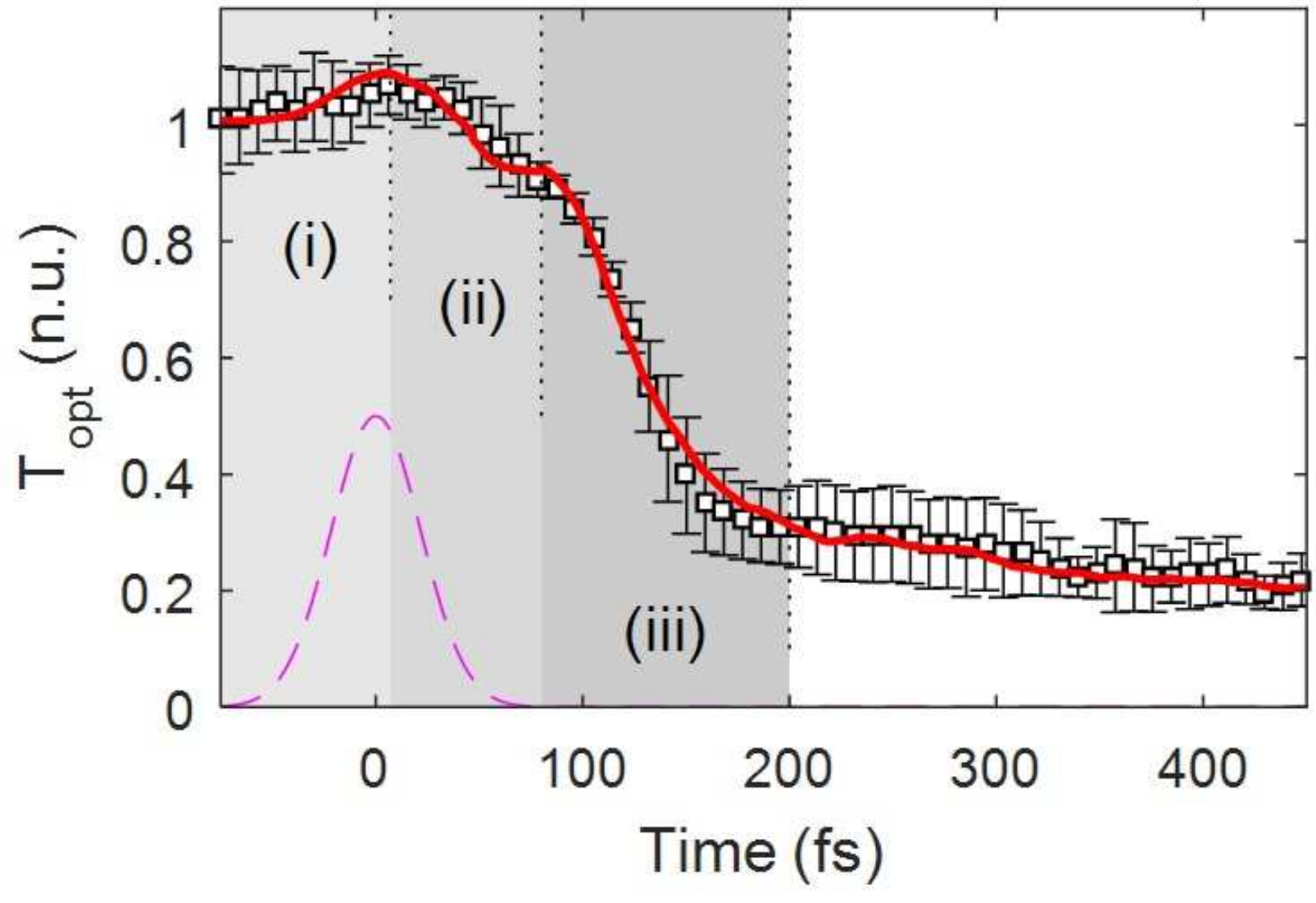}  
 \caption{\label{fig:Graphitization_transmission}  
 Calculated transmission of the optical pulse of a wavelength of 630 nm, for the average absorbed dose of 0.8 eV/atom, in the layer thickness of 38 nm (at the time instance of 400 fs; red solid line). FEL photon energy was 47.4 eV, pulse duration was 52.5 fs (FWHM, magenta dashed line).  
 It is compared to the experimental data (open black squares with errorbars).  
 Intervals (i)-(iii) denote different stages of graphitization.  
 The figure is reproduced from Ref.~\citeref{Tavella2017a}.  
 }  
\end{figure}  

All these stages can be seen in Figure~\ref{fig:Diamond_diffractions} showing atomic snapshots in diamond under irradiation with an FEL pulse with the absorbed dose of 0.8 eV/atom, photon energy of 47.4 eV, and FWHM pulse duration of 52.5 fs. One can clearly observe the formation of the graphite planes. Note that the overdense graphite planes are changing orientations at the edges for a sufficiently large simulation box: in this figure 512 atoms in the box was used, whereas for 216 atoms box formed planes were always perfectly aligned in the simulations. This  indicates a presence of multiple nucleation centers of the new phase, as in the case of the homogeneous phase transition~\cite{Lu1998,Rethfeld2002a}.  

These changes of the atomic structure are reflected in the evolution of the diffraction patterns. In Figure~\ref{fig:Diamond_diffractions}, the diamond reflections (111) are present at around 42\degree, (220) at around 74\degree, and (311) at around 91\degree at the beginning of the simulation. Then, the peaks (220) and (311) become reduced significantly during the graphitization. Later, they almost disappear. In contrast, the peak (111) is only slightly reduced. The reason for that is that this particular reflection of diamond coincides with a reflection of the overdense graphite, due to their equal density. At later timescales, when expansion of the overdense graphite takes place, the peaks shift towards smaller angles (not shown). The overlap of the diamond and graphite peaks may complicate the analysis of the experimentally recorded diffraction patterns. Luckily, as Figure~\ref{fig:Diamond_diffractions} shows, an additional peak (002) at around 50\degree emerges. It corresponds to graphite structure, clearly marking the phase transition. Additional peaks at 40\degree are observed here, corresponding to the defected structure of the bending graphite planes -- those peaks are absent in case of perfectly oriented graphite planes. 

Apart from the structure of the non-thermally created graphite phase, we can also analyze its dynamical properties. 
The vibration spectrum, Eq.(\ref{Eq:spectrum}), calculated for irradiated diamond is shown in Fig.~\ref{fig:Diamond_0.7_autocorr}(a). During the non-thermal graphitization, coherent acoustic phonon excitation takes place (see spectra at times after $\tau \sim 200-300$ fs).  
Such coherent excitation of phonons is in agreement with the model of displacive excitation of coherent phonons ($A_1$ symmetric phonons)~\cite{Zeiger1992}, and was known previously for nonthermal melting in different materials~\cite{Grigoryan2014}.  
The excited modes correspond to  $A_{2u}$ mode at $\sim 900$ cm$^{-1}$ -- the characteristic frequencies of $D_{6h}^4$ symmetric graphite~\cite{dre82:4514,mau04:075501}.  

At this time period, the optical phonon modes are inhibited in comparison to the density of states (DOS) characteristic to the equilibrium graphite~ \cite{Al-Jishi1982}. As the material tends to a new equilibrium, the energy is transferred to the optical phonons ($\tau = 700$ fs), leading to the relative intensity ratio approaching that of the equilibrium phonon spectra. This step of the diamond to graphite transition can be characterized by the onset of optical phonons at $\sim 1600$ cm$^{-1}$ for the lattice modes of $E_{2g2}$ and $E_{1u}$ symmetry~\cite{dre82:4514,mau04:075501}. This spectrum already closely resembles known graphite phonon DOS, although at high temperature (these phonons have much larger amplitude than the equilibrium ground state ones)~\cite{Al-Jishi1982}. This is reflected by the autocorrelation function, Fig.~\ref{fig:Diamond_0.7_autocorr}(b), where a sudden enhancement of vibrational amplitude $\sim{200}$ fs after the pump pulse can be seen.  
The predicted temperature increase is consistent with the increase of the atomic kinetic temperature estimated from the molecular dynamics simulation (up to $\sim 1600$ K)~\cite{Medvedev2013e}. 

This ultrafast solid-to-solid phase transition -- graphitization of diamond under an FEL pulse irradiation -- is unambiguously reflected in the evolution of the transient optical properties~\cite{Tavella2017a}. Comparison between the experimentally measured transmittance of the optical probe pulse (wavelength of 630 nm) with theoretical predictions is shown in Figure~\ref{fig:Graphitization_transmission}. The transmission curve is normalized to the initial transmission of non-irradiated diamond. It exhibits the characteristic multistep process of graphitization described above: (i) initial electronic excitation, (ii) band gap collapse, and (iii) atomic relocation~\cite{Tavella2017a}.  
The remarkable agreement between the theoretical calculations and the recent experiment confirms the transient timescales of the ultrafast graphitization, completed within $\sim 150-200$ fs -- to our knowledge, the fastest solid-solid transition observed up to now.

\subsubsection{Thermal melting of silicon}  
\label{Sec:Si_LDL}  

\begin{figure*}  
 \includegraphics[width=0.9\textwidth, trim = {30 300 30 10}]{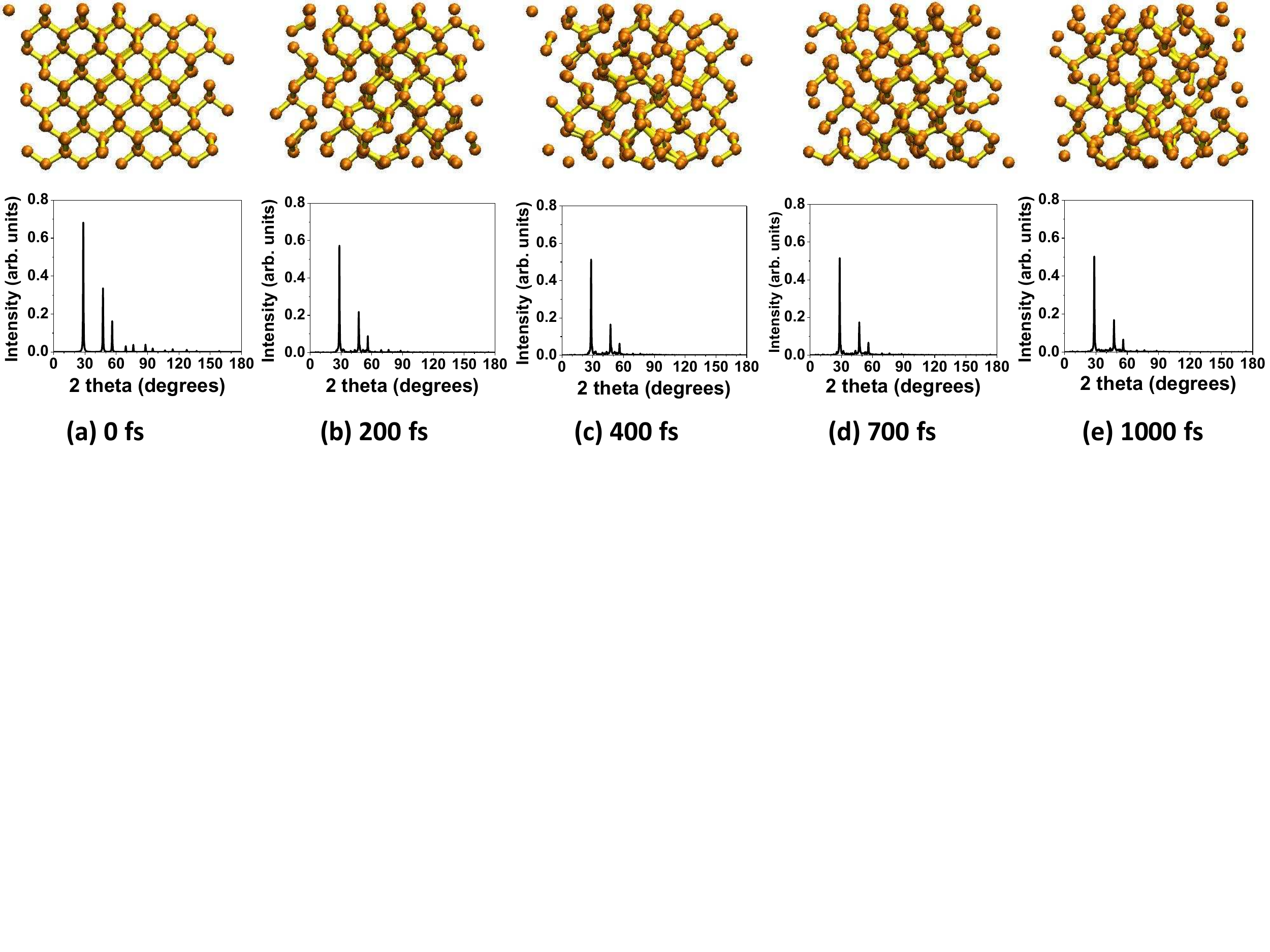}  
 \caption{\label{fig:Silicon_thermal_0.72}  
 (Top raw) Snapshots of silicon during thermal melting into low-density liquid phase at different time instants after the FEL pulse irradiation corresponding to the absorbed dose of 0.72 eV/atom, photon energy of 30 eV, and FWHM pulse duration of 60 fs. (Bottom raw) Powder diffraction patterns for the corresponding atomic structures obtained with x-ray photons of wavelength of $1.54$ \AA.  
 }  
\end{figure*}  

As we discussed in the previous papers~\cite{Medvedev2015c,Tkachenko2016,Medvedev2017}, silicon can undergo various phase transitions depending on the absorbed dose: a thermal one, occurring for the doses about $\sim 0.65$ eV/atom, and a nonthermal one, for doses above $\sim 0.9$ eV/atom. Thermal melting turns crystalline silicon into low-density liquid by heating the lattice via electron-ion coupling mechanism (nonadiabatic energy exchange discussed in section \ref{Sec:Low_En_electrons}). Nonthermal melting quickly leads to high-density liquid phase via an interplay of the lattice heating and nonthermal modification of the interatomic potential, similar to the case of diamond graphitization. The latter case will be studied below in section \ref{Sec:Si_HDL}.  

Irradiation of silicon with x-ray radiation providing an absorbed dose $0.65 - 0.9$ eV/atom induces the following processes. Firstly, high-energy electron cascades deliver photoabsorbed energy to low-energy domain via impact ionizations~\cite{Medvedev2015c}. Later, low-energy electrons couple to the ions, providing them with energy via nonadiabatic coupling, Eqs.(\ref{Eq:Boltzmann}--\ref{Eq:Heat_rate}).  
During this phase, electronic structure of the material is changing. A band gap collapse follows, indicating a transition into a semimetallic phase. Once the lattice is heated enough, the atomic structure transforms into the new phase. Figure \ref{fig:Silicon_thermal_0.72} shows the evolution of the atomic structure of silicon after irradiation with an FEL pulse of 30 eV photon energy, 60 fs FWHM duration, and 0.72 eV/atom absorbed dose. After approximately $\sim 300 - 500$ fs, one can see an onset of the melted phase. Diffraction patterns demonstrate that the short-range order is still preserved in the new state, persisting after equilibration of electronic and atomic temperatures on the timescale of $\sim 1$ ps~\cite{Medvedev2015c, Medvedev2017}.  

As the stable liquid phase of silicon is a high-density liquid, we expect that LDL phase will densify at longer timescales, unless it is quickly cooled down to be frozen in this low-density state. Presumably, transition to the HDL would take at least a few tens of picoseconds at the considered near-threshold absorbed doses~\cite{Ionin2013}. However, such timescales are too long to access with our present approach, and require dedicated investigations. 
These aspects are beyond the scope of the current discussion. 

\begin{figure*}  
 \includegraphics[width=0.9\textwidth, trim = {0 10 0 0}]{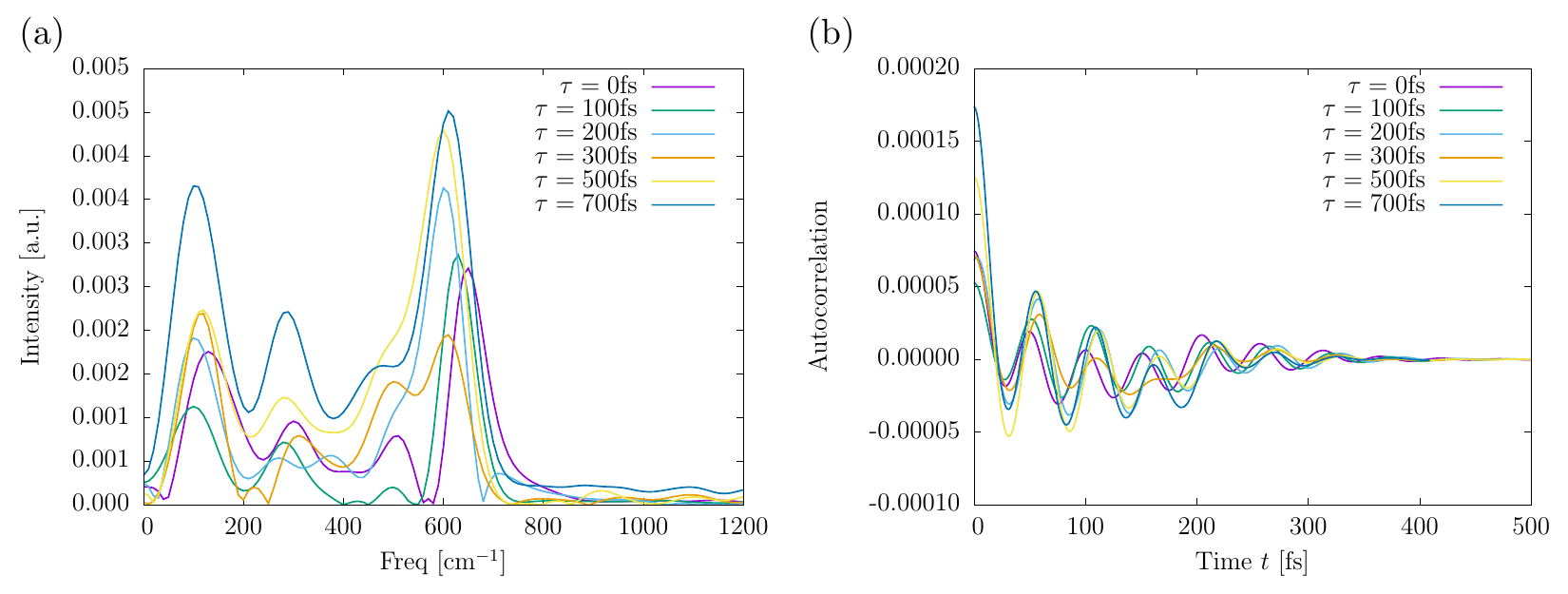}  
 \caption{\label{fig:Silicon_0.72_autocorr}  
 Calculated vibrational spectrum from the autocorrelation function of swarm of trajectories.  
 (a) The vibrational spectrum for various time delays $\tau$ after the pump pulse.  
 (b) The autocorrelation function for various time delays $\tau$, where the time $t$ is defined in Eq.~(\ref{Eq:spectrum})  
 }  
\end{figure*}  

\begin{figure}  
 \includegraphics[width=0.45\textwidth, trim = {30 20 30 30}]{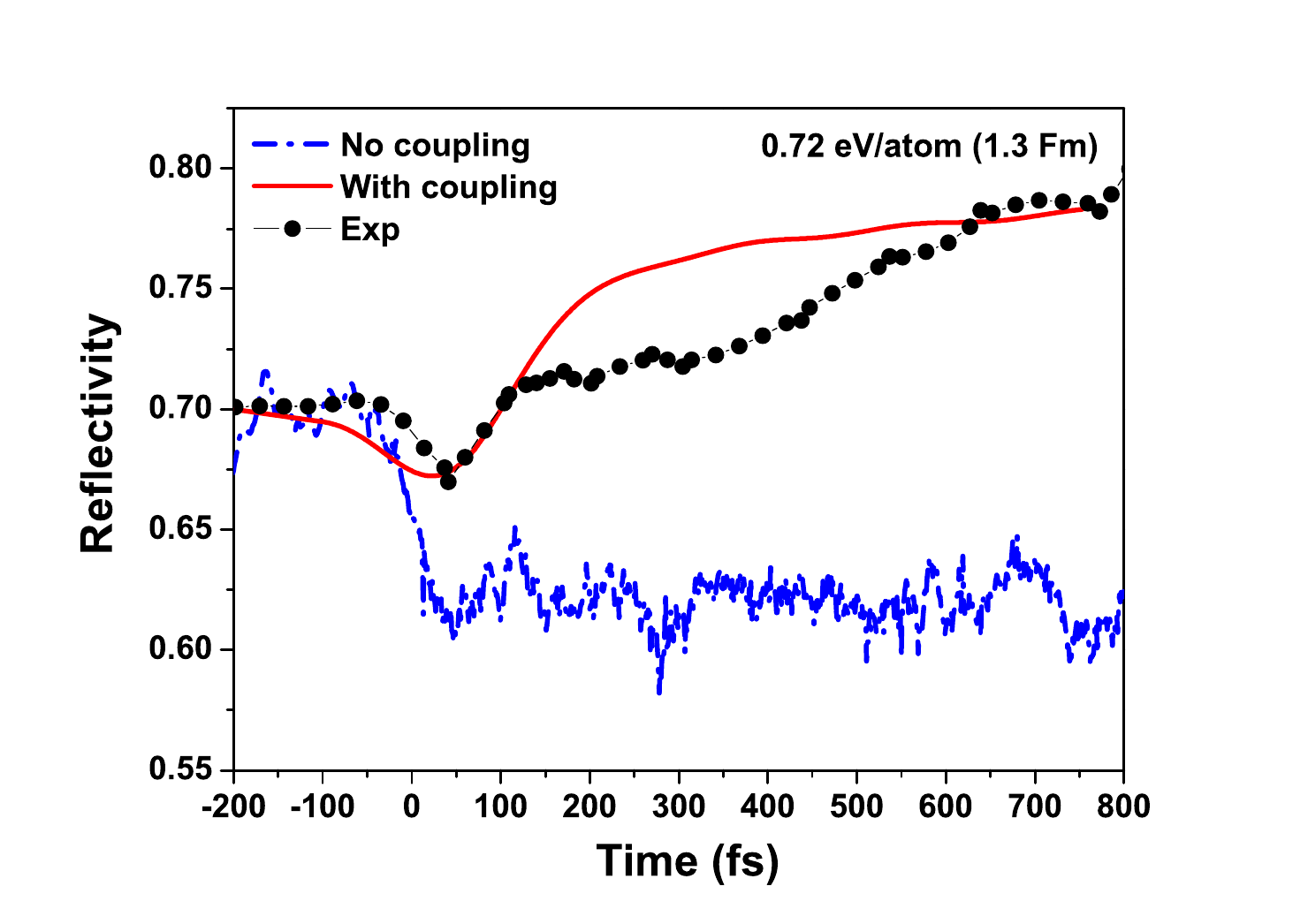} 
 \caption{\label{fig:Silicon_reflectivity_0.72}  
 Comparison of the calculated and experimental~\cite{Sokolowski-Tinten1995} reflectivity of silicon at 625 nm probe wavelength under $70.5\degree$ incidence for the fluence corresponding to 0.72 eV/atom absorbed dose. The data were convolved with 60 fs gaussian probe pulse. Two cases are compared: with electron-ion coupling, and without it (BO approximation). The figure is reproduced from Ref.~\citeref{Medvedev2017}.}  
\end{figure}  

To analyze the dynamical properties, the temporal evolution of silicon spectra is shown in Fig.~\ref{fig:Silicon_0.72_autocorr} for the case of the absorbed dose of 0.72 eV/atom. At the beginning $\tau=0$ fs, the initial phonon spectra are close to the equilibrium spectra of silicon.  
The peak around $600$ cm$^{-1}$ corresponds to the optical phonons in the vicinity of the $\Gamma$ point~\cite{cho94:2221,let07:252104}.

At later times after $\sim 300 - 500$ fs, silicon crystal enters a regime of much higher lattice vibration amplitude, which is a signature of melting. The phonon peak at around  $600$ cm$^{-1}$ is then shifting to lower frequencies at times around $300 - 500$ fs, indicating phonon softening~\cite{Hunsche1996}. It occurs due to the ongoing transition to the low-density liquid state~\cite{Medvedev2015c}. This observation is consistent with the diffraction patterns shown above in Figure~\ref{fig:Silicon_thermal_0.72}. The presence of the local order in the melted phase is also consistent with the experimentally observed low-density liquid phase created in solid silicon by FEL-irradiation~\cite{Beye2010}. The increase of the atomic temperature can be seen in the gradual increase of the amplitude of the autocorrelation function in Figure~\ref{fig:Silicon_thermal_0.72}(b).

We can compare the calculated optical properties of irradiated silicon with experimentally available data. Unfortunately, such data exists only for femtosecond optical pulses~\cite{Sokolowski-Tinten1995}. However, as we discussed in Refs.~\citeref{Tkachenko2016,Tkachenko2016a}, the electrons after VUV irradiation as well as after an optical one -- at the same near-threshold absorbed doses -- relax quickly to an equilibrium Fermi distribution. The two cases become nearly identical within a few femtoseconds after the exposure, as long as we consider bulk material with periodic boundaries, without any essential contribution of the particles and energy transport. Such a comparison is shown in Figure~\ref{fig:Silicon_reflectivity_0.72}.  

This figure also clearly shows that the observed reflectivity overshooting (i.e., the fact that final reflectivity of the irradiated sample is higher than the initial one) is a result of the thermal lattice heating. If we exclude nonadiabatic electron-ion coupling, overshooting does not show up~\cite{Medvedev2017}. These results confirm the idea presented earlier in Ref.~\citeref{Ziaja2015}: the overshooting effect observed in experimental data~\cite{Maltezopoulos2008, Gahl2008} is a consequence of the ion heating and the resulting band-gap shrinkage.  
This observation proves that it is essential to step out beyond the Born-Oppenheimer approximation when modeling the evolution of irradiated solids. Otherwise, important non-adiabatic mechanisms may be missing, and a proper description of the solid evolution can never be achieved.

\subsubsection{Coulomb explosion of \Csxty crystal layer}  
\label{Sec:C60}  

\begin{figure*} 
 \includegraphics[width=0.9\textwidth, trim = {30 150 30 20}]{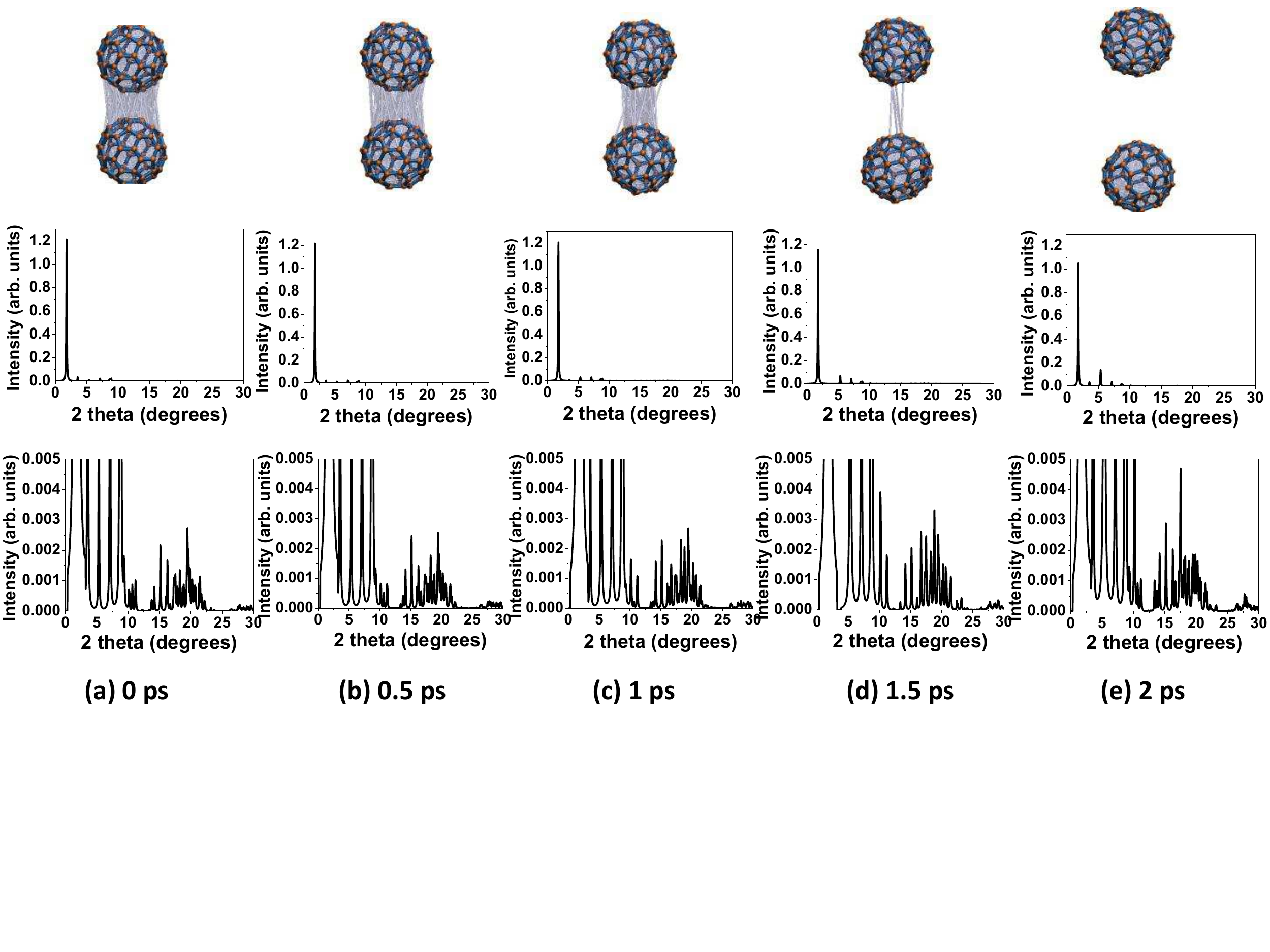}  
 \caption{\label{fig:C60_diffractions}  
 (Top raw) Snapshots of \Csxty\ crystal during Coulomb explosion at different time instants after the FEL pulse irradiation corresponding to the absorbed dose of 0.28 eV/atom, photon energy of 92 eV, and FWHM pulse duration of 30 fs. Periodic boundaries are imposed along X and Y axis, whereas there are free surfaces along Z-direction. Reproduced from Ref.~\citeref{Toufarova2017} (Middle raw) Powder diffraction patterns for the corresponding atomic structures obtained with x-ray photons of $1.54$ \AA. (Bottom raw) Zoom into the tails of the diffraction patterns indicating intermolecular connections.  
 }  
\end{figure*}  

In the recent paper, we studied behavior of thin layers of \Csxty\ crystal under irradiation with FELs x-ray pulses~\cite{Toufarova2017}.  
Our predictions indicate that irradiated \Csxty\ crystal disintegrates into single intact fullerenes. The observed fullerene behavior is caused by a Coulomb explosion induced by the charging of fullerene cages. It was confirmed by a comparison of these calculations with the calculations performed assuming all electrons to be confined within the system, which preserves charge neutrality. In the latter case, no breaking of \Csxty\ crystal was predicted~\cite{Toufarova2017}.  

The unbalanced charge is produced due to the photoabsorption by extreme ultraviolet/soft-x-ray (XUV/SXR) laser radiation and impact ionization by photo-electrons and secondary electrons. When the energy of an excited electron is above the work function of \Csxty\ (which is 7.6 eV), the electron can be emitted leaving a positive charge behind. The repulsive forces between neighboring fullerene cation radicals then decompose the molecular crystal structure, releasing fullerenes into the vacuum.  

Figure~\ref{fig:C60_diffractions} shows calculated snapshots of the sample at different times following the FLASH irradiation of \Csxty\ crystal (for the photon energy of 92 eV, 30 fs FWHM, absorbed dose of 0.28 eV/atom), Ref. \citeref{Toufarova2017}. This figure shows the scc structure. The fcc structure simulation looks nearly identical. The calculated damage threshold of fcc structure is lower only by $\sim 10 \%$ as expected from the considerations of their cohesive energies~\cite{Lu1992}. The decomposition of the \Csxty\ layer can clearly be observed. We notice that here, the \Csxty\ cages start to separate from each other on the timescale of $\sim 2$ ps, although the unbalanced charge was created within the first $\sim 100$ fs during the electron cascading. This timescales mismatch is due to a great inertia of the fullerene molecules -- such massive objects repeal each other slowly in comparison with the atomic repulsion in case of diamond and silicon phase transitions.  

In Figure~\ref{fig:C60_diffractions}, the diffraction reflections related to the intermolecular distances shift to lower angles, which is a signature of the material expansion. Although the intensity of these peaks is very low, in comparison to the intramolecular peaks at small angles, they clearly indicate a separation of \Csxty\ cages. The fact that the intramolecular peaks stay practically unchanged confirms that intact fullerens are emitted from the irradiated layer, as seen in the atomic snapshots.  

We do not perform here an autocorrelation analysis, since the van der Waals potential used~\cite{Toufarova2017} was not specifically designed to reproduce the vibrational frequencies but only to yield the correct cohesive energy and structure. 

The damage threshold for the molecular Coulomb explosion in thin layer depends on the produced unbalanced charge due to electron emission. This, in turn, depends on the FEL photon attenuation length, and the layer thickness. Thus, we cannot present here a universal damage threshold dose. For the particular photon energy studied, \hw = 92 eV, the threshold charge was estimated to be 0.0018 electrons/atom for fcc structure of \Csxty\ crystal (or 0.002 for scc \Csxty\ molecular arrangement). In this case, the absorbed dose of 0.18 eV/atom for fcc structure (or 0.21 eV/atom for scc) produces the corresponding unbalanced charge leading to breaking of bonds between \Csxty\ cages. For comparison, experimental measurement produced the damage threshold of $\sim 0.15$ eV/atom for the identical FEL pulse conditions~\cite{Toufarova2017}.

\subsection{Medium fluence}  

\subsubsection{Graphitization of amorphous carbon}  
\label{Sec:a_C_below}

\begin{figure*}  
 \includegraphics[width=0.9\textwidth, trim = {30 230 30 20}]{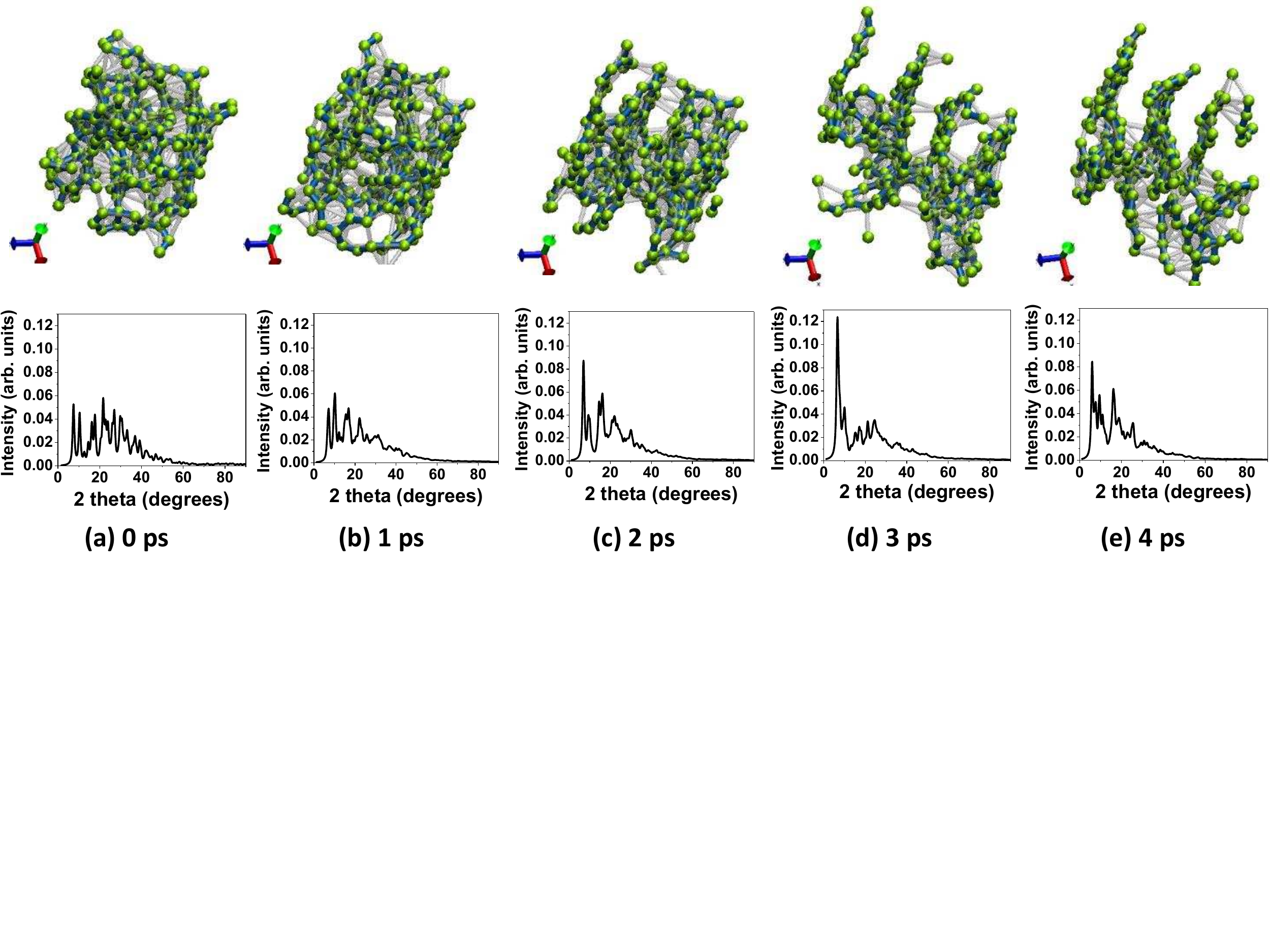}  
 \caption{\label{fig:Graphitization_aC_0.8}  
 (Top raw) Snapshots of amorphous carbon after the FEL pulse irradiation corresponding to the absorbed dose of 0.8 eV/atom, photon energy of 92 eV, and FWHM pulse duration of 30 fs. Reproduced from Ref.~\citeref{Toufarova2017}. (Bottom raw) Powder diffraction patterns for the corresponding atomic structures obtained with x-ray photons of $1.54$ \AA.  
 }  
\end{figure*}

Simulated samples of amorphous carbon (a-C) were prepared by quenching of the melted phase starting from underdense diamond (with the density equal to that of the desired a-C, $\rho = 2.6$ g/cm$^3$), heated up to the temperatures of a few thousand Kelvins. Quenching was performed by artificially setting atomic velocities to zero every few femtoseconds (similar to the standard zero-temperature MD schemes for quenching~\cite{DL_POLY}) during $\sim 5$ ps until the total energy change in the system became negligible. It indicates that a stable configuration was reached. In a series of simulations multiple initial a-C states were created and then checked for their quality. The most homogeneous and stable one was chosen for the simulation of x-ray irradiation.  

Calculations with XTANT for a-C showed that for the considered parameters of FLASH irradiation (the absorbed dose of 0.8 eV/atom, \hw = 92 eV, 30 fs FWHM), the damage threshold is $\sim 0.85-0.9$ eV/atom (cf. the experimental dose of 0.88 eV/atom~\cite{Toufarova2017}). Here, the damage threshold is defined as a dose needed to initiate spallation/disintegration of the amorphous sample, modeled with Parrinello-Rahman method for NPH ensemble. This allows to trace material expansion and its eventual fragmentation. For an above-threshold dose, the irradiated sample breaks apart into molecular fragments, and the volume of the modeled supercell expands indefinitely (see below, section~\ref{Sec:a_C_above}). For the below-damage case, no ablation was observed.  

\begin{figure*} 
 \includegraphics[width=0.9\textwidth, trim = {30 290 30 10}]{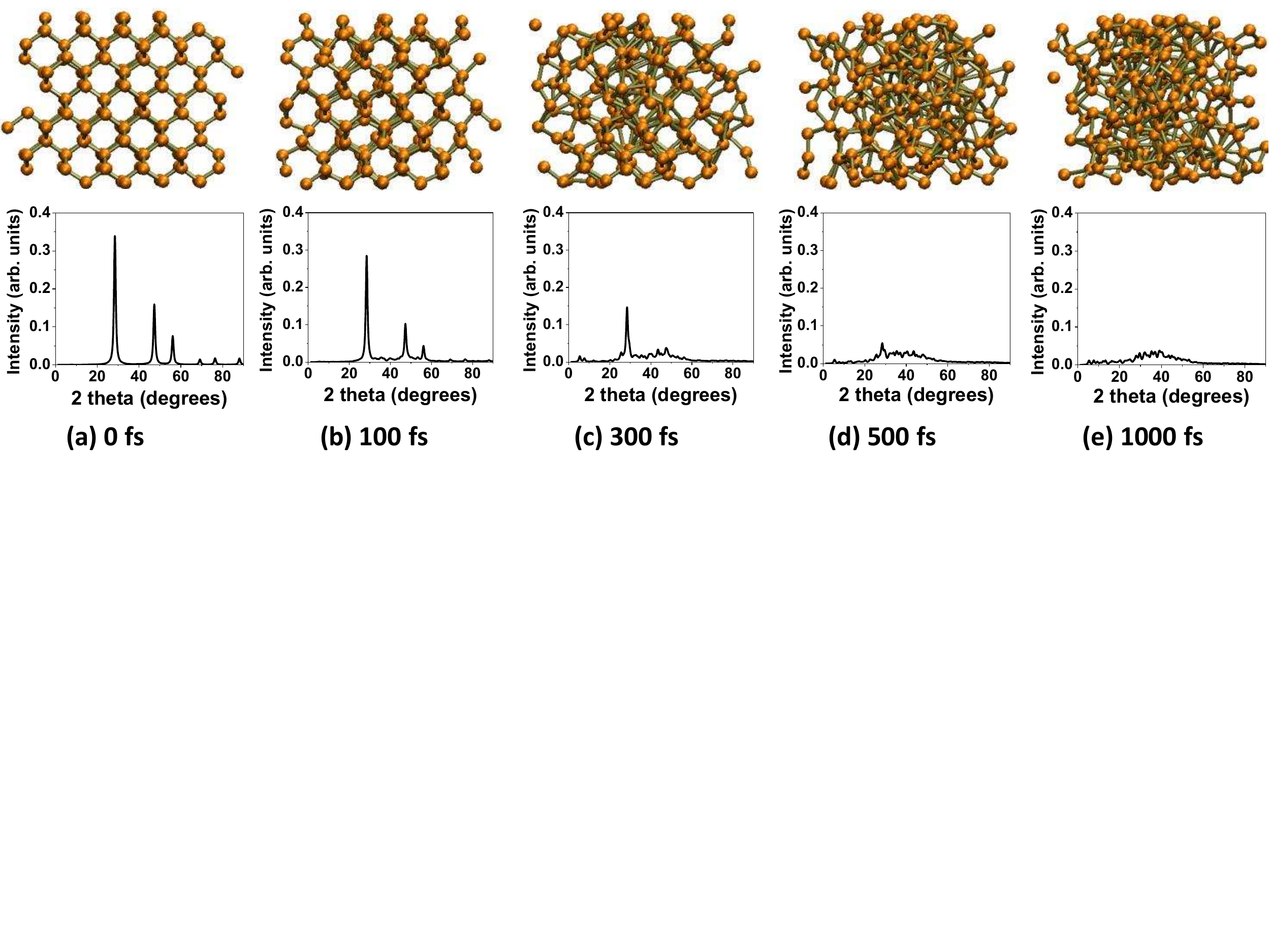}  
 \caption{\label{fig:Diffraction_Si_1.08}  
 (Top raw) Snapshots of silicon after the FEL pulse irradiation corresponding to the absorbed dose of 1.08 eV/atom, photon energy of 30 eV, and FWHM pulse duration of 60 fs. (Bottom raw) Powder diffraction patterns for the corresponding atomic structures obtained with x-ray photons of $1.54$ \AA.  
 }  
\end{figure*}  

In the below-threshold absorbed dose case ($< 0.85$ eV/atom), Figure~\ref{fig:Graphitization_aC_0.8} shows expansion of the irradiated material, which saturates after $\sim 2.5$ ps, with formation of graphite-like structures inside. As atomic snapshots in Figure~\ref{fig:Graphitization_aC_0.8} demonstrate, this process is similar to graphitization, although the formed graphite-like planes are bent and highly defected. This below-threshold expansion reproduces the experimental finding~\cite{Toufarova2017}.  

Analysis of the diffraction patterns for the corresponding atomic snapshots shows emergence of sharp peaks, indicating ordering of the material. This supports the above-mentioned scenario of graphitization of a-C under FEL irradiation.  

The good agreement of the calculated damage threshold with the experimental one confirms the reliability of the model~\cite{Toufarova2017}. To the best of our knowledge, there is no time-resolved experimental data available yet on the process of graphitization of amorphous carbon. However, we expect that the timescales of material expansion reported here are probably underestimated, as is typical for Parrinello-Rahman MD simulations.

\subsubsection{Thermal and nonthermal melting of silicon}  
\label{Sec:Si_HDL}  
%

\begin{figure*}  
 \includegraphics[width=0.9\textwidth, trim = {0 10 0 0}]{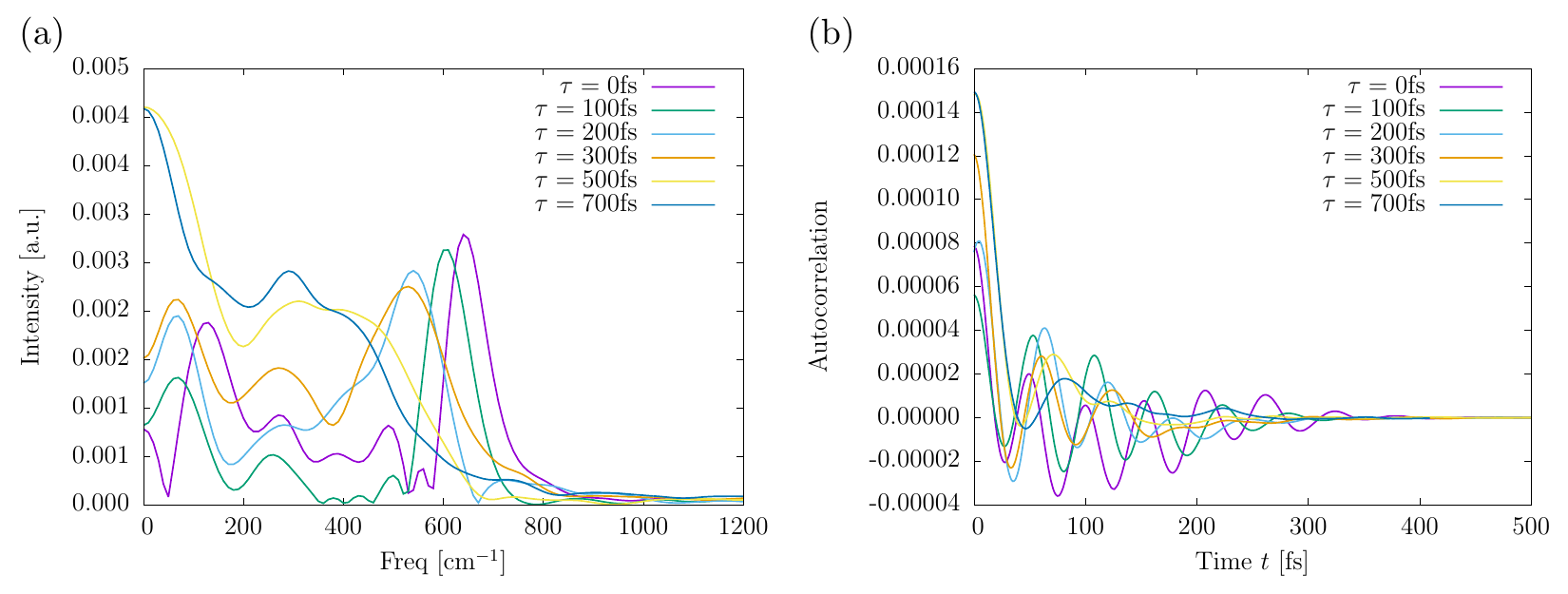}  
 \caption{\label{fig:Silicon_1.08_autocorr}  
 Calculated vibrational spectrum from the autocorrelation function of swarm of trajectories.  
 (a) The vibrational spectrum for various time delays $\tau$ after the pump pulse.  
 (b) The autocorrelation function for various time delays $\tau$, where the time $t$ is defined in Eq.~(\ref{Eq:spectrum}).
 }  
\end{figure*}  

\begin{figure}  
 \includegraphics[width=0.45\textwidth, trim = {30 20 30 30}]{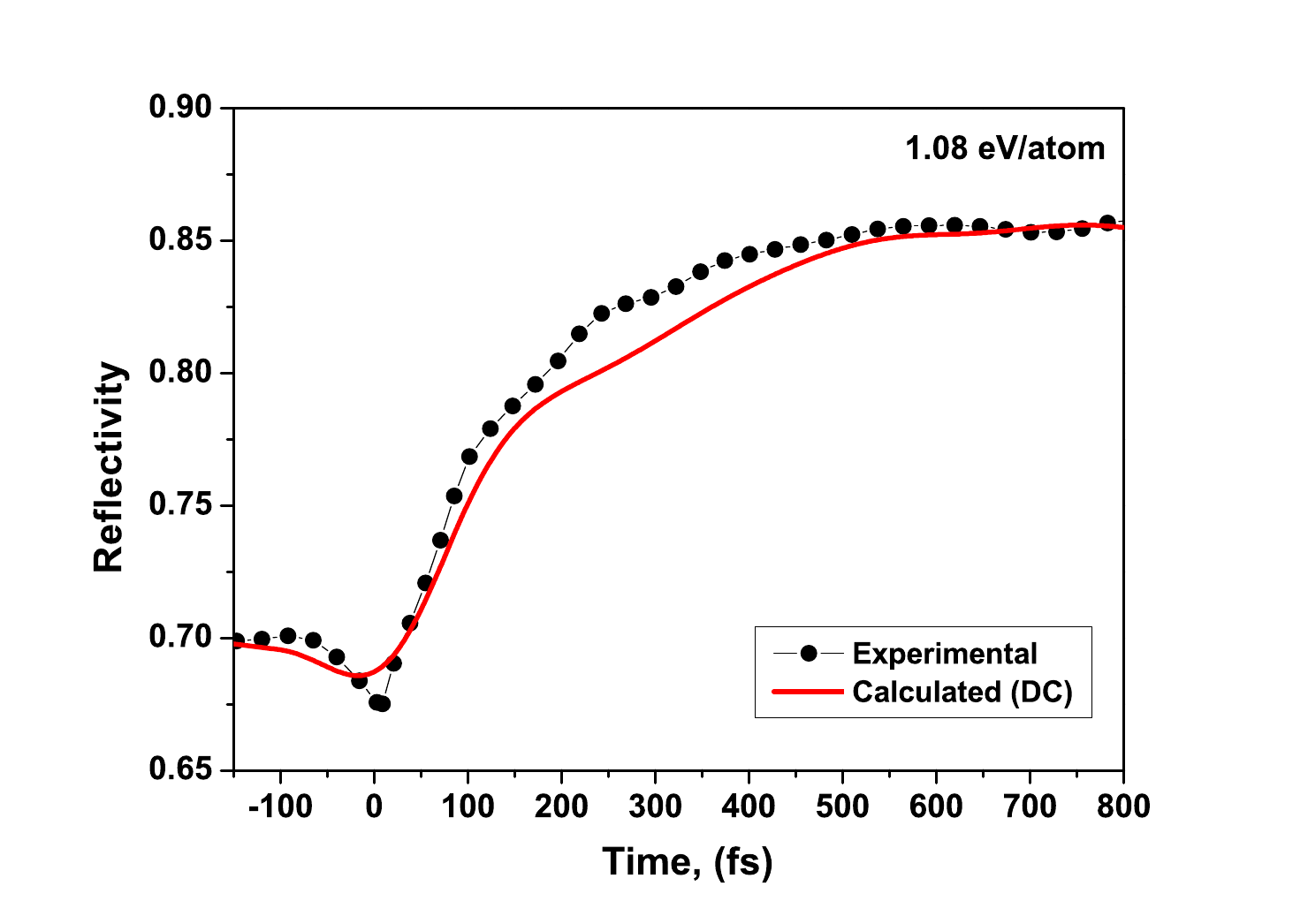}  
 \caption{\label{fig:Silicon_reflectivity_1.08}  
 Comparison of the calculated and experimental~\cite{Sokolowski-Tinten1995} reflectivity of silicon at 625 nm probe wavelength under $70.5\degree$ incidence for the fluence corresponding to 1.08 eV/atom absorbed dose. The calculated data were convolved with 60 fs gaussian probe pulse. The figure is reproduced from Ref.~\citeref{Medvedev2017}.}  
\end{figure}

We studied nonthermal melting of silicon under a femtosecond laser pulse irradiation in detail in Refs.~\citeref{Medvedev2015c, Tkachenko2016, Medvedev2017}.  
We performed the silmulations of irradiated silicon for the following FEL parameters: 1.08 eV/atom, photon energy of 30 eV, and FWHM pulse duration 60 fs. This absorbed dose is above the threshold for the nonthermal melting of $\sim 0.9$ eV/atom, calculated in~\cite{Medvedev2015c}. Note that this dose is significantly lower than the nonthermal melting threshold predicted earlier, e.g.,~\cite{Zijlstra2013}, estimated to be $\sim 2.1$ eV/atom. This is due to the fact that in our approach we included not only the changes of the potential energy surface due to excitation of electrons (Born-Oppenheimer approximation), but also the electron-ion (electron-phonon) coupling via nonadiabatic effects~\cite{Medvedev2015c}. This heating of the lattice by electrons significantly lowers the damage threshold~\cite{Medvedev2015c}.

The damage threshold can also be expressed in terms of the number of excited electrons, which is lowered from $\sim 9 \%$ within BO approximation, to $\sim 4.5 - 5 \%$ if electron-ion coupling is included. For comparison, experimental estimation of the electron density threshold is $\sim 6 \%$~\cite{Harb2008}. Thus, one can conclude that the interplay of thermal heating with the nonthermal evolution of the potential energy surface plays an important role in the damage formation in silicon~\cite{Medvedev2015c, Medvedev2017}.  

Calculations with Parrinello-Rahman MD allowed us to demonstrate that this phase transition proceeds via low-density liquid phase at a picosecond timescales into the ultimate high-density liquid phase, in a good agreement with experiments~\cite{Medvedev2015c, Beye2010}. Atomic snapshots of the material evolution are shown in Figure~\ref{fig:Diffraction_Si_1.08}. In the same figure one can also see that the material disorders on the scale of $\sim 500$ fs, with diffraction peaks almost completely disappearing. By the time of 1 ps, only the diffuse scattering background is visible in the powder diffraction.  

The evolution of vibrational spectra in silicon after absorbed dose of 1.08 eV/atom is demonstrated in Fig.~\ref{fig:Silicon_1.08_autocorr}. Again, as in the low-dose case discussed above (section \ref{Sec:Si_LDL}), at $\tau=0$ fs, the initial phonon spectra are close to the equilibrium spectra of silicon.  
After the FEL irradiation, the optical phonon peak ($\sim 600$ cm$^{-1}$) completely disappears after $\sim 500$ fs, in agreement with the conclusions drawn from the diffraction patterns. Since the optical phonons correspond to the relative motion of silicon atoms inside the primitive cell, the disappearance of the relevant peak reflects disordering of the original structure. The nonthermal transition to the high-density liquid state is thus completed~\cite{Medvedev2015c}. Again, increase of the amplitude in the autocorrelation function in Fig.~\ref{fig:Silicon_1.08_autocorr}(b) indicate heating of the lattice, however, one can also see that at times after $\sim 500$ fs disorder in the system changes the dynamics qualitatively.

The timescales of damage can be compared with experiments by tracing evolution of the optical properties. In the case of the absorbed dose above the nonthermal melting threshold, the band-gap collapse is induced via nonthermal melting~\cite{Medvedev2015c, Medvedev2017}; significant heating of the lattice is not necessary for that. For such doses, the overshooting effect allows to extract timescales of the predominant nonthermal melting. We again use the experimental data on the optical pulse irradiation~\cite{Sokolowski-Tinten1995}, as no FEL-pump time-resolved data exists as of yet to the best of our knowledge. Such a comparison is shown in  Figure~\ref{fig:Silicon_reflectivity_1.08}. One can see a very good agreement between the calculated and experimental reflectivities, indicating that the predicted timescales of damage are correct. 

\subsection{High fluence}  

\subsubsection{Spallation of amorphous carbon}  
\label{Sec:a_C_above}  

\begin{figure*}  
 \includegraphics[width=0.9\textwidth, trim = {30 250 30 0}]{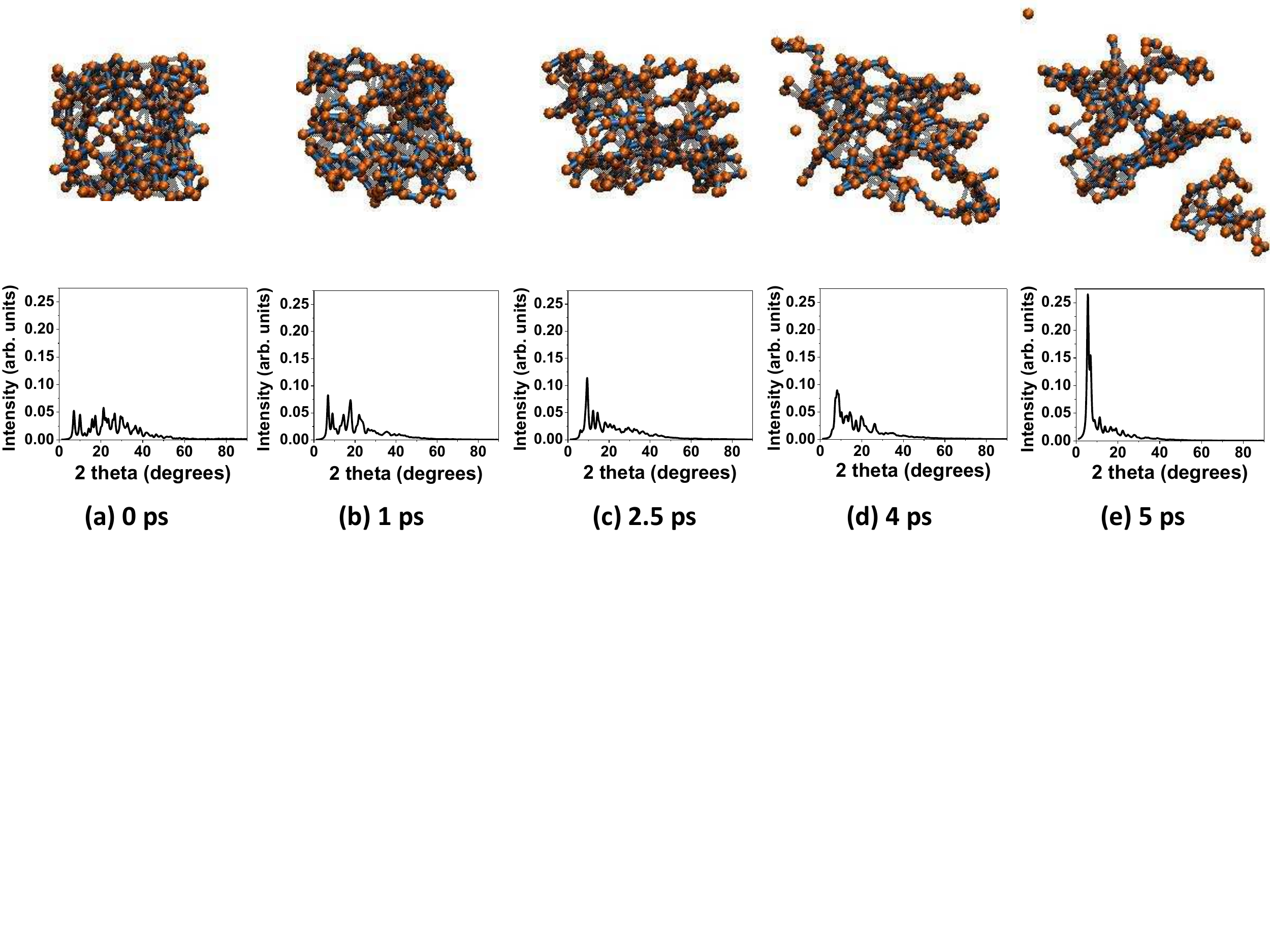}  
 \caption{\label{fig:Diffraction_aC_1}  
 (Top raw) Snapshots of amorphous carbon after the FEL pulse irradiation corresponding to the absorbed dose of 1 eV/atom, photon energy of 92 eV, and FWHM pulse duration of 30 fs. (Bottom raw) Powder diffraction patterns for the corresponding atomic structures obtained with x-ray photons of $1.54$ \AA.  
 }  
\end{figure*}  

For studying disintegration of amorphous carbon, we start with the same initial sample as discussed in section~\ref{Sec:a_C_below}. Then, we model the evolution of the sample after its irradiation with pulses of different fluences (absorbed doses) with NPH ensemble MD. After running a simulation for 10 ps, we can determine whether the super-cell volume expansion saturates or proceeds continuously and the material breaks apart. From the set of simulations, we can identify the damage threshold dose. 

The results show that after irradiation with a dose above $\sim 0.85 - 0.9$ eV/atom, a-C spallates.  
This dose corresponds to the peak density of the CB electrons of $\sim 5.5 \%$, or the electronic temperature of $\sim 13000$ K. 

The spallation proceeds via graphitization on a timescale of a few picoseconds, similar to the above-mentioned case from section \ref{Sec:a_C_below}. However, in the above-threshold case, one can clearly see spallated parts of the disintegrating sample at the time instant of 5 ps, see Figure~\ref{fig:Diffraction_aC_1}.  

Emergence of the diffraction peaks from the diffuse scattering pattern clearly shows some ordering in the structure before material disintegration. Then, material breaks apart into a few fragments, but not into atomic species, which is indicative of a spallation regime of the material removal rather than ablation~\cite{Zhigilei2006}. Note that in contrast to a thin layer of \Csxty\, here we modeled a bulk material with no open surfaces, and thus with no unbalanced charge in the system. The spallation, thus, can be considered as a thermal effect due to atomic heating via electron-ion coupling.

\subsubsection{Ablation of silicon}  
\label{Sec:Si_ablation}  

An ablation threshold of silicon was determined within the Parrinello-Rahman MD. From a set of calculations for different deposited doses, the damage threshold was estimated to be $\sim 2.6$ eV/atom. Below this threshold, Si samples demonstrated only (nonthermal) melting, without disintegration. For doses above that threshold, material disintegrated into fragments. This damage threshold dose corresponds to the maximal excited electron density of $\sim 12-13 \%$, and the electronic temperature of $\sim 20$ kK. 

\begin{figure*}  
 \includegraphics[width=0.9\textwidth, trim = {20 300 30 0}]{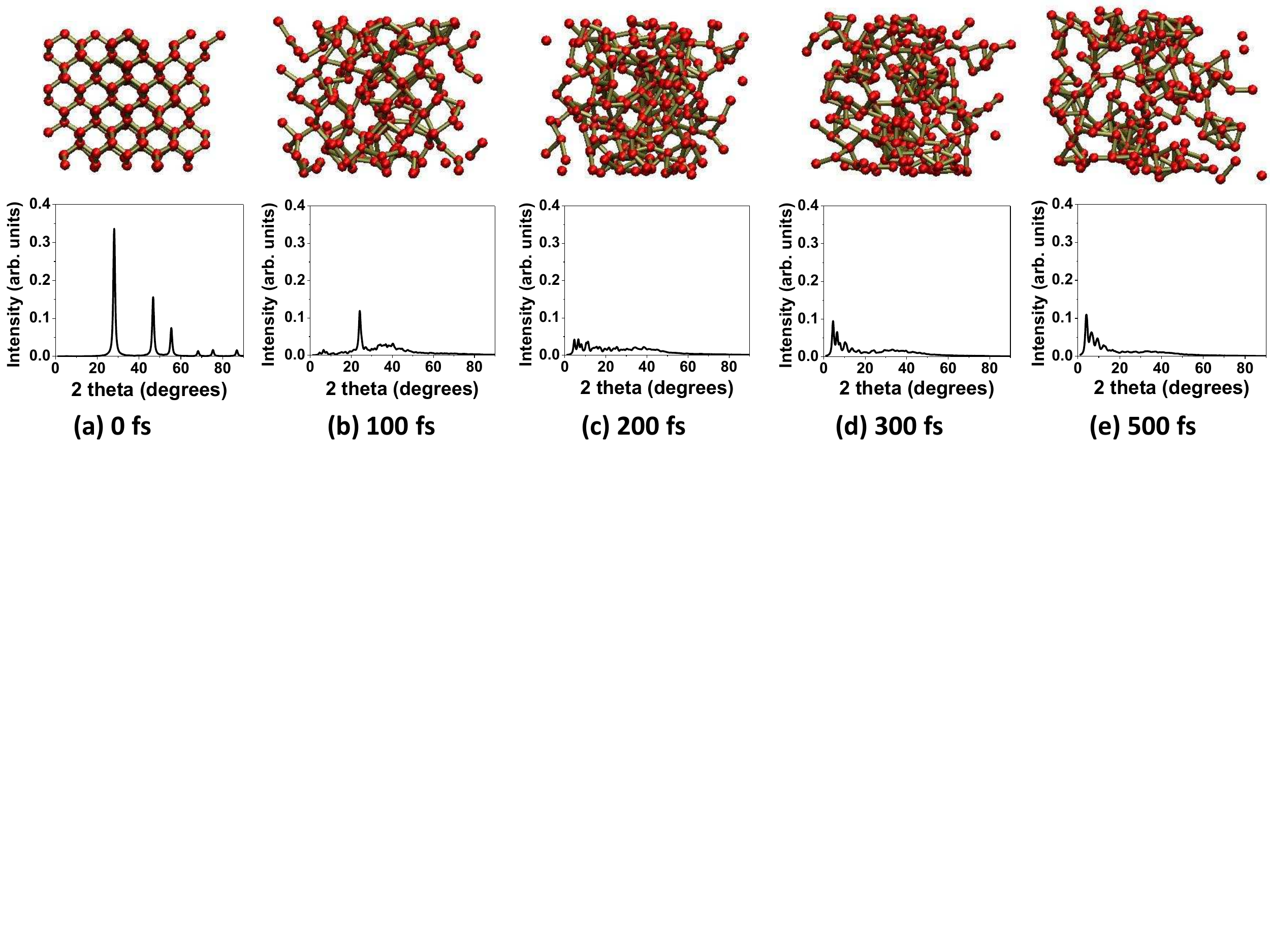}  
 \caption{\label{fig:Diffraction_Si_ablation}  
 (Top raw) Snapshots of silicon after the FEL pulse irradiation corresponding to the absorbed dose of 3 eV/atom, photon energy of 92 eV, and FWHM pulse duration of 10 fs. (Bottom raw) Powder diffraction patterns for the corresponding atomic structures obtained with x-ray photons of $1.54$ \AA.  
 }  
\end{figure*}

An example of such a simulation is shown in Fig.~\ref{fig:Diffraction_Si_ablation} for 3 eV/atom absorbed dose, delivered with an FEL pulse of 92 eV photon energy and 10 fs FWHM. The material disintegration appears to be in the ablation regime, as even individual atoms and small molecular fragments are observed.  

Silicon ablation occurs via transient nonthermal melting on subpicosecond timescales. This is consistent with the reported above nonthermal melting regime described above (section~\ref{Sec:Si_HDL}). The initial changes in the electron structure take place on sub-100 fs timescales, the band gap collapses within $\sim 40$ fs, producing semi-metallic silicon. Already within $\sim 200-300$ fs, Fig.~\ref{fig:Diffraction_Si_ablation} shows that powder diffraction reflections vanish into the rising diffuse scattering background. Thus, a complete loss of structural order takes place at this ultrashort timescale. After that, however, new diffraction peaks emerge. They confirm formation of small molecular fragments, emitted from the disintegrating bulk sample. This process is a nonthermal ablation with the contribution from thermal electron-ion heating. We note again that due to inclusion of non-adiabatic effects beyond the Born-Oppenheimer approximation, our predicted ablation threshold is significantly lower than previously predicted purely nonthermal thresholds within the BO approximation (respectively, $\sim 2.6$ eV/atom vs $\sim 4-6$ eV/atom~\cite{Jeschke2002}). This prediction should be validated by experiments.  

However, again, due to application of the Parrinello-Rahman MD scheme, we expect that the timescales of ablation are underestimated. The experimental studies would probably reveal longer times of silicon ablation.

\subsection{Ultrahigh fluence: warm dense matter formation}  
\label{Diamond_WDM}  

\begin{figure*}  
 \includegraphics[width=0.9\textwidth, trim = {30 290 30 20}]{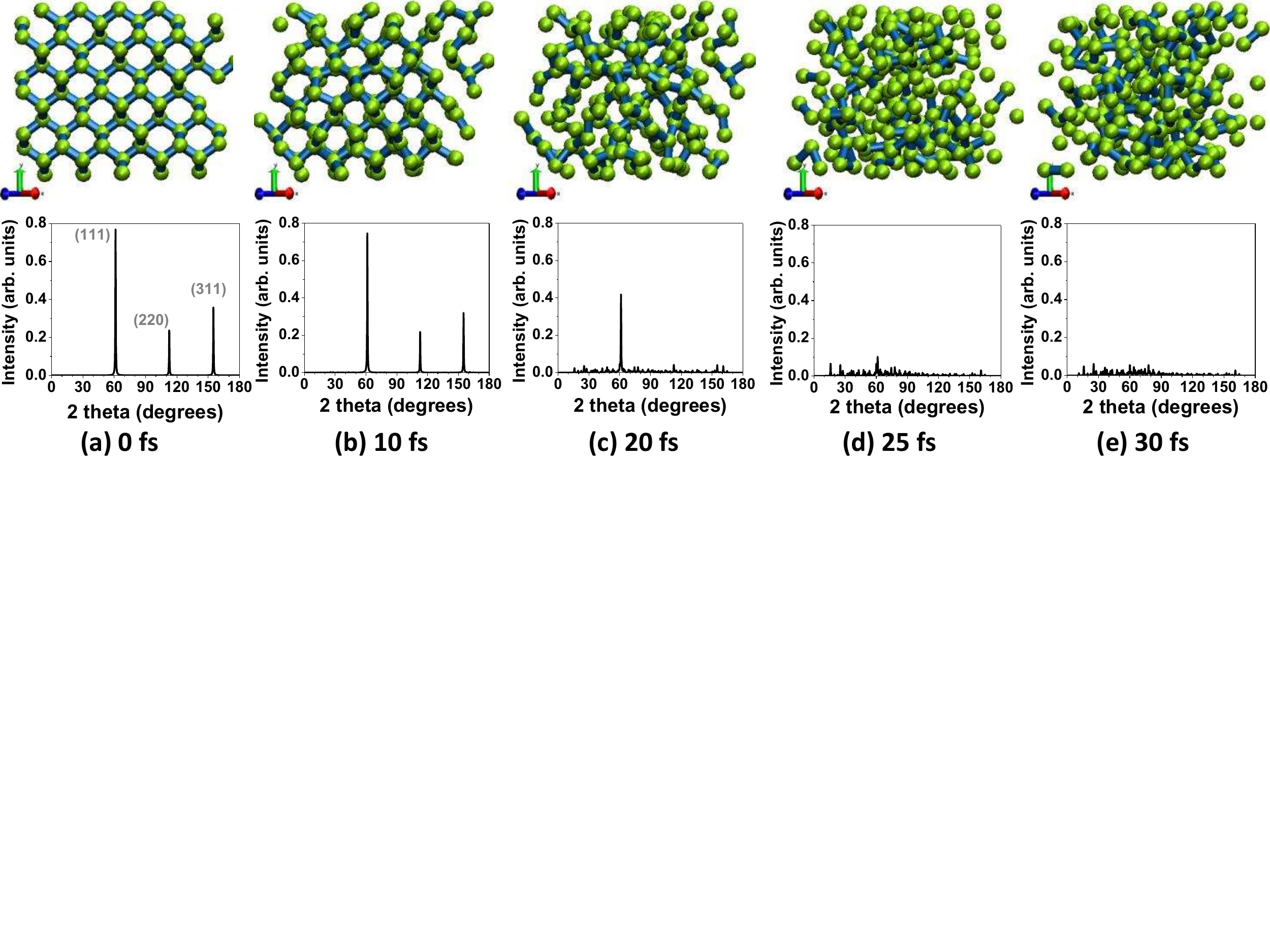}  
 \caption{\label{fig:Diamond_diffraction_18.5}  
 (Top raw) Snapshots of diamond after the FEL pulse irradiation corresponding to the absorbed dose of 18.5 eV/atom, photon energy of 6100 eV, and FWHM pulse duration of 10 fs; reproduced from Ref.~\citeref{Sci_Rep}. (Bottom raw) Powder diffraction patterns for the corresponding atomic structures obtained with x-ray photons of $1.54$ \AA.
 }  
\end{figure*}  

In order to test the limits and capabilities of our developed hybrid approach, we performed a set of simulations at very high deposited doses, leading to warm dense matter (WDM) formation, Ref.~\citeref{Sci_Rep}. The simulations of diamond irradiated with an extremely intense x-ray FEL pulses (at average absorbed doses $18.5 - 24.9$ eV/atom; 6.1 keV photon energy; 5 fs FWHM) performed with XTANT code demonstrate that the atomic structure quickly disorders, on a timescale of a few tens of femtoseconds, see atomic snapshots in Fig.~\ref{fig:Diamond_diffraction_18.5}.  
Diamond transiently undergoes through a stage similar to the graphitization discussed above (section~\ref{Sec:Nonthermal_diamond}), lasting only for a few femtoseconds. From there on, a complete disordering of the sample proceeds.

These effects are visible in the simulated diffraction patterns: the nearest neighbour peak, corresponding to the reflection (220), disappears faster than the (111) peak. That is due to the fact that the peak (111) is present in diamond as well as in overdense graphite formed before material expansion, see Fig.\ref{fig:Diamond_diffractions}. Thus, a presence of both peaks (220) and (111) indicates a diamond structure, whereas presence of the peak (111) only, with the peak (220) absent, indicates a transient overdense graphite-like state. 

This effect is confirmed by the powder diffraction patterns in Fig.~\ref{fig:Diamond_diffraction_18.5} (e.g. see the snapshot at 20 fs). After that phase, a quick atomic rearrangement follows which leads to the sample disordering at times $>20$ fs. This is due to the fact that the absorbed dose lies so much above the graphite damage threshold that it triggers atomic disordering in the graphite-like state as soon as it is formed.

This process is clearly of nonthermal nature at its early stage, i.e., until 15-20 fs. The ion temperature increases due to the nonthermal changes of the potential energy surface, and not due to the electron-ion coupling, as confirmed by a comparison with a dedicated simulation within the Born-Oppenheimer approximation. Such a simulation produced nearly identical result, showing a negligible contribution of the electron-ion coupling at such extremely short timescales~\cite{Sci_Rep}.  

The calculated diffraction peaks in Fig.~\ref{fig:Diamond_diffraction_18.5} show qualitative agreement with the recent experiment~\cite{Inoue2016}: the intensity of peak (220) decreases faster than the peak (111) both in our simulation and in experiment. However, quantitatively, the simulated diffraction peaks vanish significantly faster (by the time of $\sim 25$ fs) than the experimental ones ($> 80$ fs). This observation indicates the limit of validity of our approach reached in this particular case. 

\begin{figure*}  
 \includegraphics[width=0.9\textwidth, trim = {40 320 180 10}]{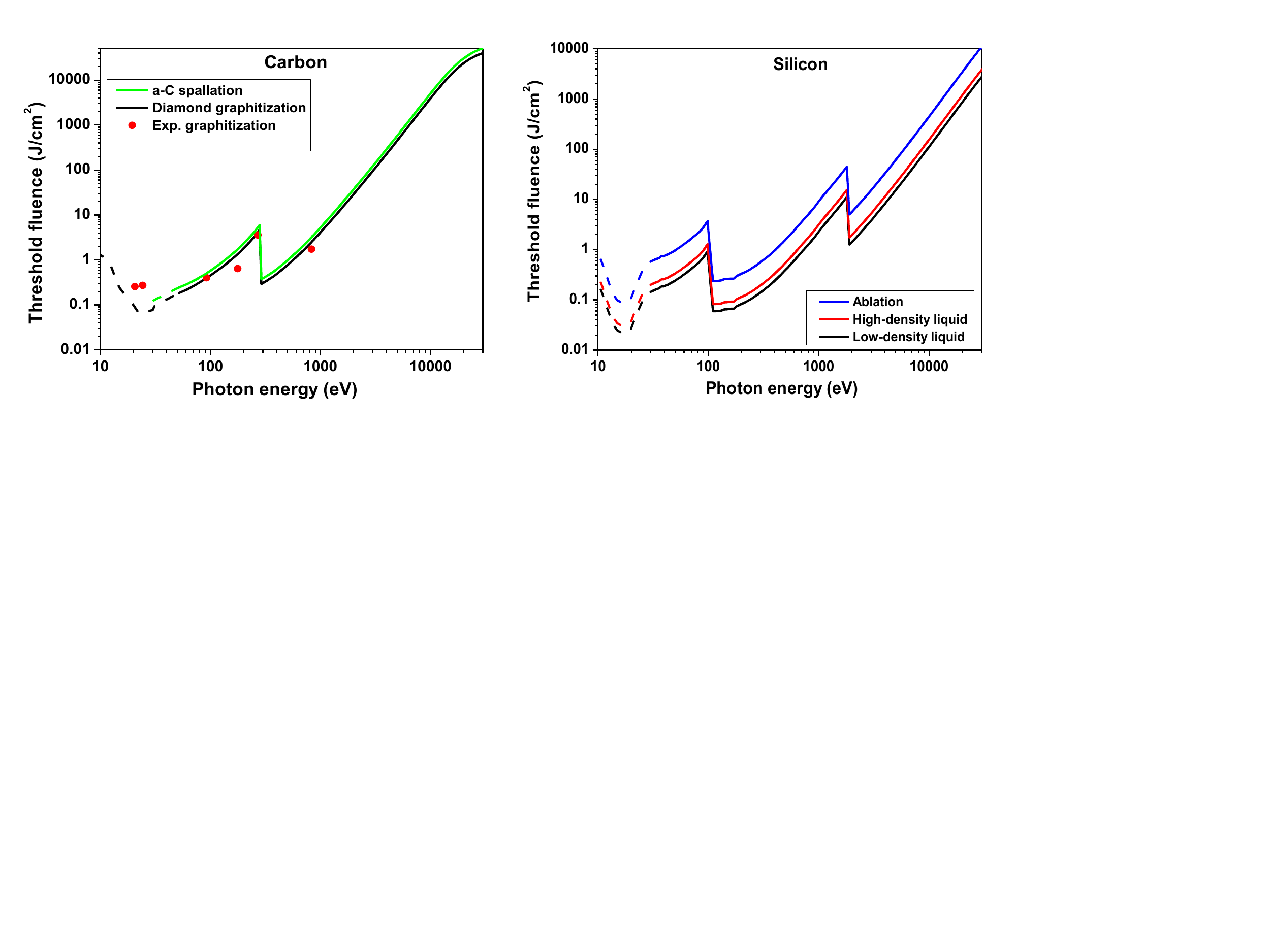}  
 \caption{\label{fig:Threshold_fluences}  
 Damage threshold fluences for: (left panel) carbon allotropes -- graphitization of diamond and spallation of amorphous carbon; and (right panel) for crystalline silicon -- melting into low-density liquid, high-density liquid, and ablation, calculated with XTANT. Experimental data for graphitization of diamond are shown for comparison; points for photon energies below 300 eV are reproduced from Ref.~\citeref{Gaudin2013}, while the point at 830 eV is from Ref.~\citeref{DastjaniFarahani2017}.  
 }  
\end{figure*}  

The shifts of diffraction peaks (111) and (220) measured in experiments were smaller than $0.15\degree$~\cite{Inoue2016}. The lack of any significant shifts of maxima positions of both Bragg peaks indicates that the material expansion due to ablation was insignificant at the experimental timescale of 80 fs. It justifies the usage of MD simulation scheme at a constant volume (V=const, NVE ensemble) which we applied. We note that application of the NPH Parrinello-Rahman scheme in this case resulted in an explosive expansion of the material at the modeled timescales of 30 fs, yielding a clear disagreement with experimental data. This, again, is in line with our argumentation above on the underestimation of the timescales within this MD scheme.

\section{Discussion}  
\label{Sec:Discussion} 

\subsection{Damage thresholds vs photon energy}  
\label{Sec:D_vs_hv}  
%

As we have discussed in the course of this paper, in many cases for materials irradiated with femtosecond x-ray pulses, there are universal damage threshold doses. Knowing the photon attenuation length in the material, one can evaluate the corresponding threshold fluence with the inverse of Eq.(\ref{Eq:Dose_fluence}). It relies upon the assumption of the negligible electron transport, so that the delivered energy is distributed in the material according to the photon penetration profile. Limitations of the model and possible improvements will be discussed in more detail in the next section.

The calculated threshold fluences for various damage channels in allotropes of carbon and in crystalline silicon are presented in Figure~\ref{fig:Threshold_fluences}. They could be used as guidance for preparation of experiments at FEL facilities, which utilize photon energies from few tens of eV to tens of keV. Note, however, that they are yet to be validated by experiments at hard x-rays.

\subsection{Limitations of the model}  
\label{Sec:limitations}  

Our results from high-fluence regime indicate a few limitations of the XTANT model. We can estimate the limiting fluence or an absorbed dose. Our results discussed in section~\ref{Diamond_WDM} showed only qualitative agreement with experiment, implying the absorbed dose of $\sim 18.9$ eV/atom is already too high for quantitative model application. On the other hand, comparison with the optical properties of diamond after irradiation with $\sim 5.4$ eV/atom produced a reasonably good agreement with experiment in Ref.~\citeref{Tkachenko2016}. Thus, the limitation of the model is somewhere in the range of $5.4 - 18.9$ eV/atom.  
A few reasons for that can be identified.  

Firstly, we can expect that the created high-energy photo-electrons are, in reality, cascading for longer times than estimated with the model. This is due to the fact that the XTANT code currently uses cross section for electron impact ionization, calculated for neutral medium with the complex dielectric function \cite{Medvedev2013e,Medvedev2015}. In case of a strong ionization of the sample by a high radiation dose deposited, the sample neutrality quickly breaks down. To the best of our knowledge, there are no rigorously derived impact ionization cross sections in highly excited solids yet available.  

In Ref.~\citeref{Sci_Rep} we showed that (artificially) reducing impact ionization cross sections slows down the diamond damage, but it is not sufficient to achieve a quantitative agreement with experiment on warm dense matter formation.  

Secondly, K-shell holes created may strongly perturb the electronic band structure, which is not taken into account in the present model. However, in Ref.~\citeref{Sci_Rep} the number of K-shell holes was shown to be small enough not to influence significantly the sample evolution. 

The most important effect which is not accounted in XTANT seems to be related to the fact that the model relies on the transferable tight binding parameterization, whose parameters were fitted to the equilibrium configurations of different carbon phases \cite{Xu1992}. This approximation misses the effect of the shifts of the electronic energy levels due to the presence of excited electrons. This problem is also known in the plasma community in the context of the ionization potential depression (IPD) \cite{Vinko2012a, Son2014, Ziaja2013a}. With the increasing temperature within the heated solid, higher charges appear within the sample (cf. Fig.~5 in Ref.~\citeref{Son2014}). The energy levels within the band correspondingly move down. This supports the expected lowering of the impact ionization rate as the ionized nuclei attract the valence electrons stronger, and the corresponding binding energies increase. 

Electrons occupying the valence levels below the Fermi level form bonding states, whereas electrons populating the levels in the conduction band above the Fermi level contribute to antibonding states. Thus, lowering of the conduction band levels beyond the Fermi level in the strongly heated diamond may temporally change the bonding from repulsive to attractive. This effect may stabilize diamond on the way to the warm dense matter state and prolong the timescales of WDM formation. Since transient amount of excited electrons in diamond is close to a metal density, perhaps similar effects to the bond hardening observed in metals can be transiently expected in diamond during warm dense matter formation~\cite{Ernstorfer2009,Grigoryan2014a}. This conjecture may be tested in the future by a dedicated ab-initio modeling, such as, e.g., finite-temperature DFT. 

Another kind of a limitation of the model is related to the periodic boundary conditions applied. In case of the normal incidence hard x-rays irradiation, when heating is homogeneous within the material down to the depth of a few microns or more, periodic boundary conditions can be applied with a good accuracy as it was done throughout the paper. For lower photon energies, or grazing incidence angles, the photon attenuation length is small and the periodic boundary conditions are inapplicable. Near the plasmon resonance, the attenuation length may be as short as a few nanometers, which induces strong gradients in the absorbed energy, and triggers strong particle and heat transport. These effects hinder the application of the model to the photon energies below about 50 eV.
For very low energies, in the ultraviolet or optical regime, additional nonlinear effects such as photoabsorption and inverse Bremsstrahlung must be taken into account~\cite{Medvedev2010d}.

And last but not least of the discussed issues, the electron-ion coupling scheme which we currently apply in XTANT seems to have limitations. In our attempt to apply it to GaAs, the calculated nonadiabatic electron-ion coupling rate seemed to underestimate the experimental values extracted from the optical probe data~\cite{Tkachenko2017}. The reason for this discrepancy is likely that GaAs has a very narrow conduction band minimum. For its proper sampling, one would need a large number of k-points~\cite{Liu2017}. Thus, our electron-ion energy exchange, so far written for the Gamma-point (Eq.(\ref{Eq:Landau})), must be extended to include multiple $k$-points. This will be a topic of a separate study.

\subsection{Future development}  
\label{Sec:XMOLECULE}  

The reliability of the XTANT code, confirmed by its quantitative agreement with various experimental results, proofs the correctness of the approach and demonstrates powerful capabilities of hybrid models in general. Consequently, one can think of further improvements of the model in order to extend its applicability. 

Firstly, in order to enable the usage of the model for other materials, a non-orthogonal TB parameterizations can be implemented. As was mentioned, e.g., in Ref.~\citeref{Mehl1996}, a non-orthogonal TB allows for easier construction of transferable parameterizations. It is already available for many elemental solids~\cite{Mehl1996}. 

Secondly, in order to go beyond the simple periodic boundary conditions, one can incorporate particle and energy transport effects into the model. One way is to include additional source and sink terms into the electronic and atomic equations. For example, a proper tracing of the electronic transport can be done by extending the MC module and the rate-equation/thermodynamics module to account for spatial dimensions, see, e.g.,~\cite{Lipp2017}. In this way, a sample can be discretized into a set of simulation boxes. Each of them would have its specific absorbed dose. Exchange of particles and energies between each other could also be accounted for.

As shown, e.g., in Ref.~\cite{Rethfeld2002}, at low fluences the thermalization of low-energy electrons may take long times. In order to take this effect into account, one could replace the rate-equations and the TTM with a proper Boltzmann electron-electron collision integrals~\cite{Rethfeld2002}, or a full Boltzmann transport equation~\cite{Ziaja2008}. As nonequilibrium electron distribution couples differently to the ions, this may affect the atomic dynamics in the case of long electron thermalization times. Nonequilibrium electron distribution could also affect optical coefficients (cf. Eq.(\ref{Eq:rpa})), which may improve their agreement with experiments (e.g. in Fig.~\ref{fig:Silicon_reflectivity_0.72}).

Thirdly, the calculations presented in Sections \ref{Sec:a_C_above} - \ref{Diamond_WDM} show that the high-fluence regime lies on the border of applicability of the tight binding method. The latter relies on the ground-state parametrization and therefore should be applied with care under such conditions. In order to reliably describe irradiation at higher fluences, one would need to replace the tight binding module with a more suitable approach. Since at higher material excitations we expect to reach states far from equilibrium, we require a robust method not relying on the ground-state approximation, which casts doubts on the applicability of the DFT schemes. 

In Ref.~\citeref{Hao2015}, an {\em \textit{ab-initio}} scheme XMOLECULE for the calculation of the electronic structure in molecular ensembles within the Hartree-Fock (HF) approximation was developed. It includes multiple-hole configurations of molecules or solids formed during XFEL pulses and is able to provide necessary information about the electronic states, also including the influence of core holes on the interatomic forces under strongly nonequilibrium conditions. Incorporation of such a method into XTANT would allow to simulate previously unreachable conditions, such as x-ray-generated warm or even hot dense matter.  

A disadvantage of such an approach is its high computational costs. In order to decrease them, one could try to use XMOLECULE to obtain classical interatomic force-fields with the parameters adjusted on-the-fly. One of the fitting methods suitable for such an implementation is the force-matching method~\cite{ercolessi1994interatomic}. Another option would be to create a HF-based TB parameterizations on-the-fly, similarly to the density-functional-based tight binding (DFTB) methods~\cite{Porezag1995,Elstner1998}. 


\section{Summary}  

We proposed a hybrid model specifically designed to treat femtosecond free-electron-laser irradiation of solids. The corresponding code XTANT has been used here for carbon- and silicon-based materials. It provides a good description of various damage channels: nonthermal graphitization of diamond and amorphous carbon (a-C), thermal spallation of a-C, molecular Coulomb explosion of \Csxty\ crystal, thermal melting of silicon into low-density liquid phase, nonthermal melting of silicon into high-density liquid phase, ablation of silicon. It even gives some insights into warm dense matter formation in diamond. 

Wherever available, a comparison with time-resolved experimental data was provided. This was done by extracting experimental observables from XTANT, such as transient transmission and reflectivity coefficients for an optical probe pulse. The reliability of the XTANT code, confirmed by its quantitative agreement with various experimental results, proves the correctness of the approach and demonstrates powerful capabilities of hybrid models in general. 

For each studied case, we presented transient atomic snapshots and powder diffraction patterns. The time-resolved diffraction patterns can be used for a comparison with future x-ray-pump x-ray-probe experiments at FELs. Damage thresholds calculated for a wide range of photon energies were also shown. 

Finally, we presented a discussion on the limitations of the developed model and outlined directions for its future improvements. 

\section{Conflict of interest}  
The authors declare no competing interests.

\section{Acknowledgments}  
The authors thank Jerome Gaudin, Ichiro Inoue, Harald Jeschke, Libor Juha, Robin Santra, Ryszard Sobierajski, Sang-Kil Son, Franz Tavella, Oriol Vendrell, Ilme Schlichting, Sven Toleikis for valuable discussions.  
Z. Li thanks the Volkswagen Foundation for partial financial support through Peter Paul Ewald-Fellowship.  
Partial financial support from the Czech Ministry of Education (Grants LG15013 and LM2015083) is acknowledged by N. Medvedev.

\section{References}

\bibliographystyle{unsrt}
\bibliography{My_Collection}

\end{document}